\PassOptionsToPackage{table}{xcolor}
\documentclass[sigconf]{acmart}

\usepackage[english]{babel}
\usepackage[nobiblatex]{xurl}
\usepackage{tikz}
\usepackage{amsmath}
\usepackage{mathtools}
\usepackage{stfloats} 
\usepackage{enumitem}
\usepackage{booktabs}
\usepackage[T1]{fontenc}
\setlist[itemize]{leftmargin=20pt}
\usetikzlibrary{fit}
\usetikzlibrary{calc}
\usetikzlibrary{shapes.misc, positioning}
\usetikzlibrary{automata,arrows,positioning,calc}
\usepackage{adjustbox}
\definecolor{LIGHTGREEN}{HTML}{C1CC99}
\definecolor{DARKPURPLE}{HTML}{3c096c}

\definecolor{BLUE}{HTML}{06AED5}
\definecolor{DARKBLUE}{HTML}{05668D}
\definecolor{LIGHTRED}{HTML}{B98389}
\definecolor{PINK}{HTML}{F2BAC9}
\usepackage{balance} 
\definecolor{RED}{HTML}{DA3E52}
\definecolor{YELLOW}{HTML}{F8BE57}
\definecolor{VIOLET}{HTML}{7E6B8F}
\definecolor{PURPLE}{HTML}{B118C8}
\definecolor{ORANGE}{HTML}{FF784F}
\newcommand\T[1]{\textbf{#1} }

\renewcommand\footnotetextcopyrightpermission[1]{} 
\setcopyright{none}

\newcommand{\greyrule}{\arrayrulecolor{black!30}\midrule\arrayrulecolor{black}}


\begin{document}
\title{Ethereum's Proposer-Builder Separation:\\ Promises and Realities}

\author{Lioba Heimbach}
\affiliation{ETH Zurich \country{Switzerland}}
\email{hlioba@ethz.ch}
\author{Lucianna Kiffer}
\affiliation{ETH Zurich \country{Switzerland}}
\email{lkiffer@ethz.ch}
\author{Christof Ferreira Torres}
\affiliation{ETH Zurich \country{Switzerland}}
\email{christof.torres@inf.ethz.ch}
\author{Roger Wattenhofer}
\affiliation{ETH Zurich \country{Switzerland}}
\email{wattenhofer@ethz.ch}

\begin{abstract}
With Ethereum's transition from \textit{Proof-of-Work} to \textit{Proof-of-Stake} in September 2022 came another paradigm shift, the \textit{Proposer-Builder Separation (PBS)} scheme. PBS was introduced to decouple the roles of selecting and ordering transactions in a block (i.e., the \textit{builder}), from those validating its contents and proposing the block to the network as the new head of the blockchain (i.e., the \textit{proposer}). In this landscape, proposers are the validators in the Proof-of-Stake consensus protocol, while now relying on specialized block builders for creating blocks with the highest value for the proposer. Additionally, \textit{relays} act as mediators between builders and proposers. We study PBS adoption and show that the current landscape exhibits significant centralization amongst the builders and relays. Further, we explore whether PBS effectively achieves its intended objectives of enabling hobbyist validators to maximize block profitability and preventing censorship. Our findings reveal that although PBS grants validators the opportunity to access optimized and competitive blocks, it tends to stimulate censorship rather than reduce it. Additionally, we demonstrate that relays do not consistently uphold their commitments and may prove unreliable. Specifically, proposers do not always receive the complete promised value, and the censorship or filtering capabilities pledged by relays exhibit significant gaps.
\end{abstract}

\maketitle
\pagestyle{plain}

\section{Introduction}

Ethereum's~\cite{wood2014ethereum} original design assigns block proposers (miners or validators) the task of building and proposing blocks to the network. Thus, block proposers must gather transactions from the gossip network and combine them into a block before broadcasting the block to their peers. Optimal profit block building is essentially the bin packing problem  which has long been known to be NP-hard~\cite{karp1972computers}, as there is limited block space and the fees received from transactions can vary widely. In practice, however, proposers have simply ordered transactions according to their \textit{gas price}, i.e., the fees paid per unit of \textit{gas}, which reflects the amount of computation required by the transaction. This has changed with the rise of \textit{decentralized finance (DeFi)} on the Ethereum blockchain. 

DeFi refers to a collection of \textit{smart contracts}, which are executable codes stored on the blockchain~\cite{buterin2014next}, that offer financial services similar to those in traditional finance, such as exchanges and lending. These smart contracts are typically transaction order dependent, meaning that the outcome of a set of transactions is determined by their order. As a result, the concept of \textit{miner/maximal extractable value (MEV)} has emerged on the Ethereum blockchain. MEV quantifies the potential revenue that can be extracted by including, excluding, or reordering transactions~\cite{daian2020flash}. 

The widespread adoption of DeFi and the related rise of MEV has led to increased complexity in block building, leading to block proposers running their own complicated strategies to maximize the profit they make. In response to this trend, and in anticipation of the transition of the Ethereum blockchain from \textit{Proof-of-Work (PoW)} to \textit{Proof-of-Stake (PoS)}, the concept of \textit{Proposer-Builder Separation (PBS)} was introduced~\cite{buterin2021proposer}. 
PBS separates block building from block proposal: instead of having the block's proposer (a PoS validator that is known in advance) perform both tasks, specialized builders are responsible for creating blocks. Builders and validators both connect to new entities called relays, which receive new blocks from builders and forward them to the validators. When it is their turn to propose a block, the validator simply picks the most profitable block and proposes it to the Ethereum network. 

With \textit{``the merge''}, which marked the end of PoW and the beginning of PoS on Ethereum, PBS was launched and its adoption has been on the rise. In this study, we analyze the current state of the PBS scheme during its opt-in phase and evaluate how well it aligns with its intended objectives.

\T{Contributions.} We summarize our main contributions in the following.

\begin{itemize}
    \item  We examine whether PBS successfully achieves its goal of decentralizing block validation by ensuring that large entities do not have an advantage in block building. We find that professionalized builders have a distinct advantage in building profitable blocks. However, PBS effectively provides all validators, regardless of size, access to competitive blocks, thus preventing hobbyists from being outcompeted by institutional players who can optimize block profitability better.
    \item  We further investigate the extent to which PBS prevents censorship, a critical design goal of the scheme. Unfortunately, our analysis reveals that PBS falls short of its goal in practice. In fact, we consistently observe that transactions from sanctioned addresses are twice as likely to be included in non-PBS blocks compared to PBS blocks. This suggests that the current implementation of PBS does not effectively prevent censorship.

\end{itemize}

Our study highlights the current challenges and open problems associated with PBS, such as builder and relay centralization, 
the role of PBS in aiding censorship, and the reliability and trustworthiness of relays. By examining these issues, we strive to help the understanding of PBS and provide insights that can direct future improvements in the design and implementation of this new scheme.

\section{Background}
\label{sec:background}
\subsection{The Ethereum Ecosystem}

\T{Ethereum Proof-of-Stake.} Since the \textit{merge} on 15 September 2022~\cite{ethereum-merge}, Ethereum has been running two layers: the execution layer and the consensus layer. The execution layer is largely similar to the former PoW protocol and is responsible for verifying and executing transactions. The consensus layer, built on top of the Beacon chain, is responsible for reaching consensus amongst \textit{validators}, i.e., the PoS equivalent of miners. We note that to become a validator one must \textit{stake}, or lock-in, 32 ETH to incentivize honest behavior. 

\begin{figure}[t]\vspace{-4pt}
\centering
\begin{tikzpicture}[scale=0.85]
\fill [black] (-0.8,0.5)circle [radius=2pt];
\fill [black] (-0.6,0.5) circle [radius=2pt];
\fill [black] (-0.4,0.5)circle [radius=2pt];
\node[fit={(0,0) (1.2,1)}, inner sep=0pt, draw=black, thick] (a) {};
\draw[<->,thick] (1.5,1.2) -- node[above] {epoch} (6.6,1.2);
\draw[-,thick] (1.35,-0.4) -- node[below] {} (1.35,1.4);

\draw[<->,thick] (1.5,-0.2) -- node[below] {slot} (2.7,-0.2);

\node[fit={(1.5,0) (2.7,1)}, inner sep=0pt, draw=black, thick] (b) {};
\node[fit={(3,0) (4.2,1)}, inner sep=0pt, draw=black, thick] (c) {};
\fill [black] (4.6,0.5)circle [radius=2pt];

\fill [black] (4.8,0.5)circle [radius=2pt];
\fill [black] (5.0,0.5)circle [radius=2pt];

\node[fit={(5.4,0) (6.6,1)}, inner sep=0pt, draw=black, thick] (d) {};
\draw[-,thick] (6.75,-0.4) -- node[below] {} (6.75,1.4);

\node[fit={(6.9,0) (8.1,1)}, inner sep=0pt, draw=black, thick] (e) {};
\fill [black] (8.5,0.5)circle [radius=2pt];
\fill [black] (8.7,0.5) circle [radius=2pt];
\fill [black] (8.9,0.5)circle [radius=2pt];
\end{tikzpicture}
\caption{Every Beacon chain slot is 12s long, and groups of 32 slots are referred to as epochs. A validator is randomly selected as a proposer for every slot to propose a block during that slot.}\label{fig:pos}\vspace{-6pt}
\end{figure}
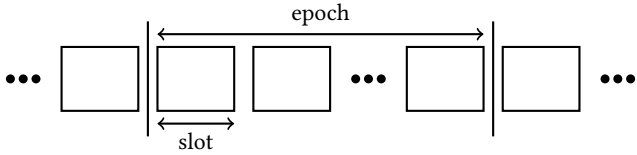

On the Beacon chain, time is split into 12 second \textit{slots}, with a group of 32 slots making up an \textit{epoch} (cf. Figure~\ref{fig:pos}). For each slot, a random validator is selected as the block \textit{proposer} along with a \textit{committee} (a group of validators in charge of verifying the newly proposed block). We note that block proposers and committees are announced  at least an epoch ahead of time, i.e.,  6.4 minutes. There is a chance for a single block to be added to the Ethereum chain in every Beacon slot. When a proposer successfully proposes a block, i.e., the block is added to the canonical chain, they receive a block reward in the Beacon chain ($\sim$0.034 ETH). The committee members partaking in the validation of the block also receive a smaller reward ($\sim$0.0000125 ETH). 

Importantly, a proposer not only receives the Beacon chain reward for proposing a block but also receives part of the block's transaction fees, as well as potentially direct payments from users as a bribe for including their transaction (totaling on average 0.1126 ETH per block). We discuss these further in the reward data of Section~\ref{sec:ethereumblockchain}. 

\T{The Ethereum Network.} Ethereum's execution and consensus layers run over two separate P2P overlays, one primarily responsible for propagating transactions and the other for consensus messages (new blocks and validator votes), respectively. Transactions sent through the P2P network are aggregated in a node's \textit{mempool} of pending transactions until they are included in a block. Validators connect directly to these networks, choosing which transactions to include in the next block from the mempool. Additionally, large validators often offer private pathways for users to send transactions to be included in a block bypassing the public mempool. Private pathways also existed pre-merge by large mining pools~\cite{ethermine}, through hidden transaction auction platforms like Flashbots \cite{flashbots}, and via private services like those offered by bloXroute~\cite{klarman2018bloxroute}. 

\subsection{Proposer-Builder Separation} 

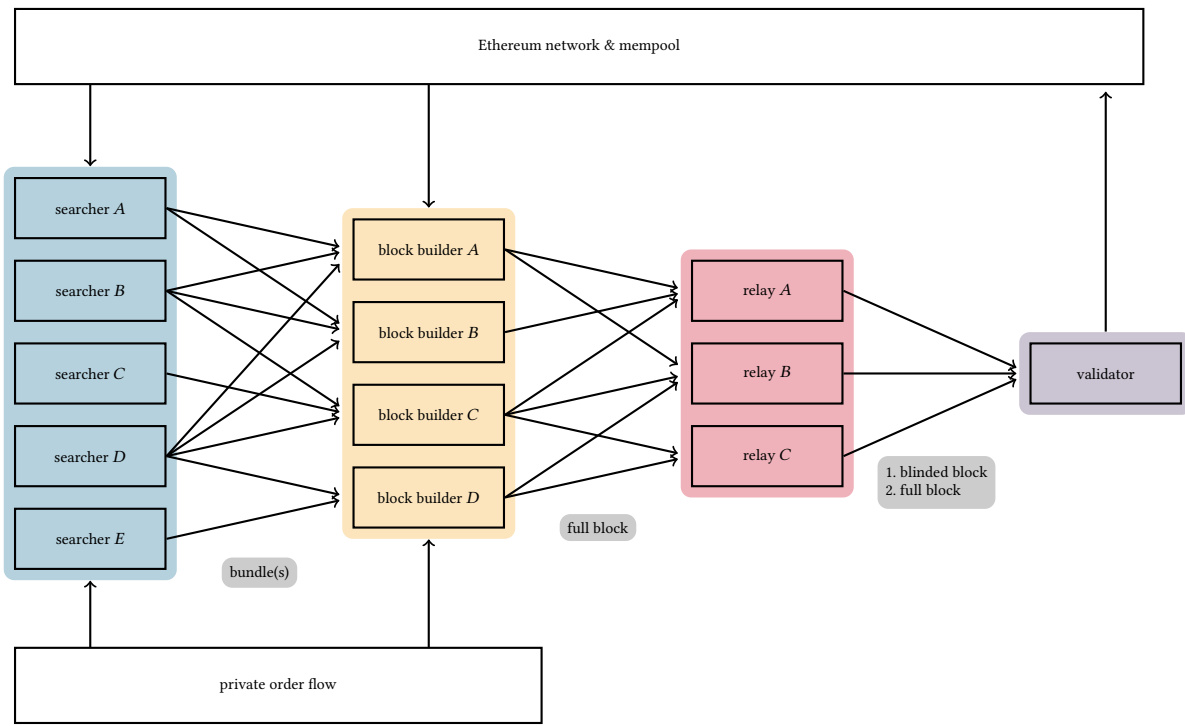
\begin{figure*}[t]
\centering
\begin{tikzpicture}[scale=1,font=\scriptsize]
    \node (rect) at (2,3.8) [draw,thick,minimum width=15cm,minimum height=1cm] (E) {Ethereum network \& mempool};

    \node (rect) at (-2,-4.7) [draw,thick,minimum width=7cm,minimum height=1cm] (E) {private order flow};

    \node [fill = DARKBLUE!30, rounded corners, minimum width =2.3cm, minimum height =5.5cm] at (-4.5,-0.55) (S1) {};
 
    \node (rect) at (-4.5,1.65) [draw,thick,minimum width=2cm,minimum height=0.8cm] (SA) {searcher $A$};
    \node (rect) at (-4.5,0.55) [draw,thick,minimum width=2cm,minimum height=0.8cm] (SB) {searcher $B$};
    \node (rect) at (-4.5,-0.55) [draw,thick,minimum width=2cm,minimum height=0.8cm] (SC) {searcher $C$};
    \node (rect) at (-4.5,-1.65) [draw,thick,minimum width=2cm,minimum height=0.8cm] (SD) {searcher $D$};
    \node (rect) at (-4.5,-2.75) [draw,thick,minimum width=2cm,minimum height=0.8cm] (SE) {searcher $E$};

    \node [fill = gray!40, rounded corners] at (-2.25,-3.2) {bundle(s)};

    \node [fill = YELLOW!40, rounded corners, minimum width =2.3cm, minimum height =4.4cm] at (0,-0.55) (B1) {};

    \node (rect) at (0,1.1) [draw,thick,minimum width=2cm,minimum height=0.8cm] (BA) {block builder $A$};
    \node (rect) at (0,0) [draw,thick,minimum width=2cm,minimum height=0.8cm] (BB) {block builder $B$};
    \node (rect) at (0,-1.1) [draw,thick,minimum width=2cm,minimum height=0.8cm] (BC) {block builder $C$};
    \node (rect) at (0,-2.2) [draw,thick,minimum width=2cm,minimum height=0.8cm] (BD) {block builder $D$};

    \node [fill = RED!40, rounded corners, minimum width =2.3cm, minimum height =3.3 cm] at (4.5,-0.55) (R1) {};
    
    \node [fill = gray!40, rounded corners] at (2.25,-2.6) {full block};
    
    \node (rect) at (4.5,0.55) [draw,thick,minimum width=2cm,minimum height=0.8cm] (RA) {relay $A$};
    \node (rect) at (4.5,-0.55) [draw,thick,minimum width=2cm,minimum height=0.8cm] (RB) {relay $B$};
    \node (rect) at (4.5,-1.65) [draw,thick,minimum width=2cm,minimum height=0.8cm] (RC) {relay $C$};

    \node [fill = gray!40, rounded corners] at (6.75,-2) {\begin{tabular}{@{}l@{}}
        1. blinded block\\
        2. full block\\
    \end{tabular}};

    \node [fill = VIOLET!40, rounded corners, minimum width =2.3cm, minimum height =1.1 cm] at (9,-0.55) (V1) {};

    \node (rect) at (9,-0.55) [draw,thick,minimum width=2cm,minimum height=0.8cm] (V) {validator};

    \draw [->,thick] (-4.5,3.3) -- ($(S1.north)$);
    \draw [->,thick] (0,3.3) -- ($(B1.north)$);
    \draw [<-,thick] (9,3.2) -- ($(V1.north)$);

    \draw [->,thick] (-4.5,-4.2) -- ($(S1.south)$);
    \draw [->,thick] (0,-4.2) -- ($(B1.south)$);

    \draw [->,thick] (SA.east) -- ($(SA.east)!0.93!(BA.west)$);
    \draw [->,thick] (SA.east) -- ($(SA.east)!0.93!(BB.west)$);
    \draw [->,thick] (SB.east) -- ($(SB.east)!0.93!(BA.west)$);
    \draw [->,thick] (SB.east) -- ($(SB.east)!0.93!(BB.west)$);
    \draw [->,thick] (SB.east) -- ($(SB.east)!0.93!(BC.west)$);
    \draw [->,thick] (SC.east) -- ($(SC.east)!0.93!(BC.west)$);
    \draw [->,thick] (SD.east) -- ($(SD.east)!0.93!(BA.west)$);
    \draw [->,thick] (SD.east) -- ($(SD.east)!0.93!(BB.west)$);
    \draw [->,thick] (SD.east) -- ($(SD.east)!0.93!(BC.west)$);
    \draw [->,thick] (SD.east) -- ($(SD.east)!0.93!(BD.west)$);
    \draw [->,thick] (SE.east) -- ($(SE.east)!0.93!(BD.west)$);
    
    \draw [->,thick] (BA.east) -- ($(BA.east)!0.93!(RA.west)$);
    \draw [->,thick] (BA.east) -- ($(BA.east)!0.93!(RB.west)$);
    \draw [->,thick] (BB.east) -- ($(BB.east)!0.93!(RA.west)$);
    \draw [->,thick] (BC.east) -- ($(BC.east)!0.93!(RA.west)$);
    \draw [->,thick] (BC.east) -- ($(BC.east)!0.93!(RB.west)$);
    \draw [->,thick] (BC.east) -- ($(BC.east)!0.93!(RC.west)$);
    \draw [->,thick] (BD.east) -- ($(BD.east)!0.93!(RB.west)$);
    \draw [->,thick] (BD.east) -- ($(BD.east)!0.93!(RC.west)$);

    \draw [->,thick] (RA.east) -- ($(RA.east)!0.93!(V.west)$);
    \draw [->,thick] (RB.east) -- ($(RB.east)!0.93!(V.west)$);
    \draw [->,thick] (RC.east) -- ($(RC.east)!0.93!(V.west)$);
    \end{tikzpicture}

\caption{Illustration of the PBS scheme. To maintain pre-trade privacy, \textcolor{DARKBLUE}{searchers} send their private transactions along with potential mempool transactions to \textcolor{YELLOW}{block builders} in bundles. The block builders then use these bundles, their own transactions, and mempool transactions to construct the most profitable block possible, which they send in full to \textcolor{RED}{relays}. The header of the most profitable block received by a relay is then sent to the \textcolor{VIOLET}{validator}. Once the validator signs the header and returns it to the relay, the full block is sent to the validator. Finally, the validator proposes the block to the consensus layer.}\label{fig:pbs}
\end{figure*}

PBS was designed for Ethereum PoS and has gained increased popularity after the merge (cf. Figure~\ref{fig:pbs}). With the PBS scheme, the former role of the proposer is separated into two new roles --- block building and block proposing. Succinctly, with PBS, block builders are responsible for creating blocks and offering them to the block proposer in each slot. The block proposer then chooses the most profitable block and proposes the block to the network. The PBS design goals are two-fold: (1) prevent censorship, and (2) decentralize transaction validation by not giving large entities an advantage in block building \cite{2022pbs,2023mevflashbots}.  

We illustrate the PBS ecosystem in Figure~\ref{fig:pbs} and detail the scheme in the following, by discussing the roles of the involved players.  The execution/consensus layers are represented as the \textit{Ethereum network and mempool} in Figure~\ref{fig:pbs}.

\T{Searchers} are Ethereum users who prioritize privacy and prefer to use a private transaction pool instead of the public mempool. Common examples of searchers include MEV bots, such as arbitrageurs, sandwich attackers, or liquidation bots, as well as Ethereum users seeking front-running protection, such as DEX traders.
With PBS, searchers send bundles containing their own transactions and possibly other transactions from the Ethereum mempool to one or multiple block builders (as shown in Figure~\ref{fig:pbs}). Searchers bid for block inclusion using either the gas price or direct ETH transfers to the address of the block builder.

\T{Block builders} receive bundles from searchers in the PBS system. Using these bundles along with their private transactions and transactions from the public mempool, block builders attempt to create the most profitable block possible. Once a block is constructed, it is sent to one or more relays in its entirety.

\T{Relays} are responsible for holding blocks from builders in escrow for validators. Their role includes accepting blocks from builders, sending the header of the most profitable block to validators, and then sending the full block to validators only after receiving a signed header. Importantly, relays keep the contents of the block private until the validator commits to proposing it for inclusion by signing the block's header.

\T{Validators} are still responsible for proposing blocks to the Ethereum network. They have the option to connect to several relays and then select the most profitable block --- the block with the highest profit for the validator --- for inclusion in the blockchain. To receive bids from the relays, a validator must install the MEV-Boost client \cite{mev-boost}\footnote{Notably, MEV-Boost is maintained by Flashbots and began as part of Flashbots' MEV transaction auction platform pre-merge.}  and add the relays from which they wish to receive bids to the config file.

For blocks built through PBS, the builder's address is listed in the block's transaction fee recipient field. Additionally, any direct payments from searchers are sent to the address listed as the block's fee recipient, i.e., the builder's address. In the block's last transaction, the builder address transfers ETH to the proposer's fee recipient address. The value of this transfer should correspond to the value agreed upon. 

\section{Data Collection}
\label{sec:datacollection}
    To measure the PBS ecosystem, we collect data from three data sources detailed below: Ethereum blockchain (cf. Section~\ref{sec:ethereumblockchain}), Ethereum network (cf. Section~\ref{sec:ethereumnetwork}), and individual data directly gathered from relays (cf. Section~\ref{sec:relaydata}). For all three, our data ranges from block 15,537,394 to block 16,950,602, i.e., from the first block after the merge on 15 September 2022 to the last block on 31 March 2023. \tableautorefname{} \ref{tab:data_collection} provides an overview of the different datasets we collected during our data collection process. It includes the number of entries and a description of the data source. We further note that we provide our aggregate data set on GitHub~\cite{projectgit}.

\begin{table}\vspace{6pt}
    \centering

    \begin{adjustbox}{width=\columnwidth}
    \begin{tabular}{@{}l r l l@{}}
       \toprule
       \textbf{Dataset} & \textbf{Entries}&\textbf{Type} & \textbf{Source}\\
       \midrule
       Ethereum blockchain  &  \begin{tabular}{@{}r@{}} 1,413,209 \\ 210,695,337 \\ 465,863,321 \\ 1,033,519,365   \end{tabular} & \begin{tabular}{@{}l@{}}blocks \\ transactions \\ logs \\ traces \end{tabular}& Erigon node \\
       \greyrule
       MEV labels   &  \begin{tabular}{@{}r@{}} 1,389,814 \\ 2,155,838 \\  1,196,662  \end{tabular}&  tx labels   &   \begin{tabular}{@{}l@{}}\url{eigenphi.io}, \url{etherscan.io} \\ \url{zeromev.org} \\ modified version of~\cite{weintraub2022flash}  \end{tabular} \\
       \greyrule     
       mempool data & 910,577,701 & tx arrival times & \url{mempool.guru} \\
       \greyrule
       relay data & 427,443,787 & proposed blocks & \url{etherscan.io}, \tableautorefname{} \ref{tab:relay_list}  \\
       \greyrule   
       OFAC & 134 & addresses & \url{treasury.gov} \\
       \bottomrule
    \end{tabular}
    \end{adjustbox}\vspace{2mm}
    \caption{Overview of the individual datasets that were collected.}
    \label{tab:data_collection}\vspace{-12pt} 
\end{table}

\subsection{Ethereum Blockchain}\label{sec:ethereumblockchain}
    We extract three types of data from the Ethereum blockchain as described below: reward data, maximal extractable value (MEV), and sanctioned address usage. To collect Ethereum blockchain data, we run an Erigon~\cite{2022erigon} execution client along with a Lighthouse~\cite{2022Lighthouse} consensus client to gather all mainnet blocks and transactions. We collect additional MEV data from third parties as described below.
    
\T{Reward Data.}  In this work we are interested in understanding the profit generated from building a block, and where this profit is going (i.e., how much is split between the builder and proposer). Though block proposers/validators also receive rewards for their work in the Beacon network, we omit these from our analysis as they are set values and orthogonal to the PBS scheme. We focus, instead, on the user-generated rewards. Note that throughout we focus on on-chain payments and do not consider possible off-chain payments. 

Since the adoption of EIP-1559~\cite{eip1559}, a user pays a fee of $f=b+p$ for their transaction consisting of a \textit{base fee} $b$ and a \textit{priority fee} $p$. The base fee is paid per unit of \textit{gas} (a measure of the computation required for a transaction) and is set by the network, adjusting dynamically according to network demand, i.e., higher demand, higher base fee. This base fee is \textit{burned} by the system and thus is not part of the block rewards. The priority fee, on the other hand, is set by the user and works as a \textit{tip} for the block creator to include the transaction. In each block, the \textit{transaction fee recipient} field is set by the creator of the block (i.e., the builder in PBS) and is the address that receives the priority fees. 

\begin{figure}
        \centering
        \includegraphics[scale =1]{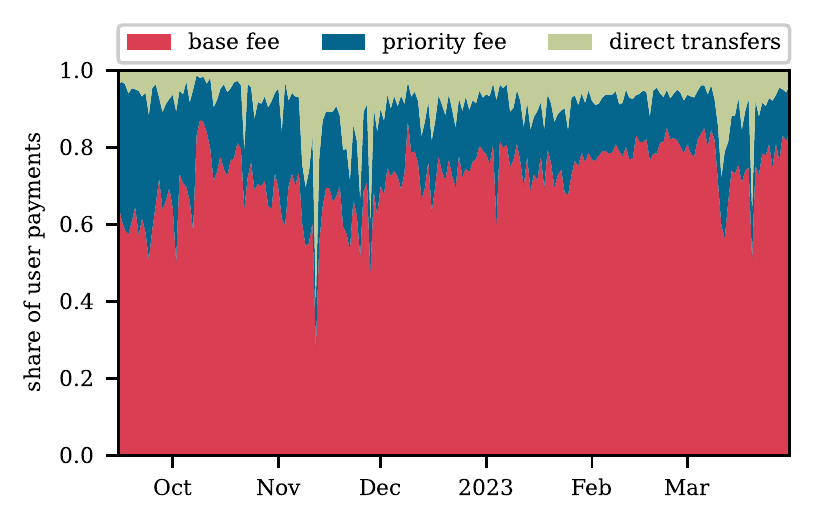}\vspace{-6pt}   
        \caption{Share of daily user payments that are burned (base fee), and those that go to the fee recipient address (priority fee + direct transfers).}
        \label{fig:feecomposition}\vspace{-4pt}          
\end{figure}

Recall that users can also ``tip'' the block creator via direct transfers to the fee recipient address. To capture these, we trace each transaction and record any internal transfer to the fee recipient address that takes place within the execution of a transaction. We record the sum of all priority fees and direct transfers as the \textit{total profit of each block}. Figure~\ref{fig:feecomposition} shows the breakdown of the user payments for each block. We observe that the majority of user fees are part of the base fee (on average, 72.3\% of user fees are burned each day). The priority fee tends to be the second largest component of user payments (a daily average of 18.4\%), while direct transfers make up the smallest part of user payments. Together the priority fee and the direct transfers make up the block's value for the proposers and builders. Thus, they (are expected to) try to maximize this. 

As a standard in the PBS scheme, the block builder pays out the proposer some amount of the profit (generally the majority) in a direct transfer as the last transaction in the block. We record this value as the \textit{proposer profit} for each block, and the difference between this value and the block profit as the \textit{builder profit}. For non-PBS blocks, the proposer profit is equal to the total value of the block.

\T{Maximal Extractable Value.} 
MEV refers to any value that can be extracted from block production beyond the standard block reward and gas fees. MEV extraction is performed by either including, censoring, or changing the order of transactions within a block. While searchers, builders, and validators have the capability to extract MEV, there is a clear hierarchy of power where searchers depend on builders to include and not ignore or steal their MEV bundles. Similarly, builders depend on validators to include these MEV bundles in their blocks and not include the validator's own bundles. The three most common types of MEV are sandwich attacks, arbitrage, and liquidations~\cite{ethereum_mev}. In short, the attacker makes a financial gain with a sandwich attack by front- and back-running the victim's trade on a DEX. Arbitrage takes advantage of price differences across DEXes for profit. Finally, liquidations close positions on lending protocols that are close to becoming undercollateralized. Whether MEV benefits or harms the Ethereum ecosystem is actively debated. Regardless, it currently accounts for a significant proportion of block rewards. 

We leverage MEV data to understand what impact MEV has after the merge as a source of income for all the parties involved in PBS, and also to understand if relays that claim to be against MEV, truly do not include MEV extraction within their forwarded blocks.
Weintraub et al.~\cite{weintraub2022flash} publicly released their scripts that detect MEV in the form of sandwiches, arbitrage, and liquidations. There also exist numerous platforms that continuously monitor newly produced blocks to identify and classify MEV. These platforms and scripts have been developed independently from each other and with different focuses.
We combine MEV data (i.e., take the union) from three different sources: EigenPhi~\cite{eigenphi}, ZeroMev~\cite{zeromev}, and our own data using a modified version of the scripts of Weintraub et al.~\cite{weintraub2022flash} to craft a large MEV dataset. It is difficult to detect MEV, the challenge is in identifying MEV patterns but once a new pattern is identified it is unlikely that flagged transactions are false positives. Thus, to have maximum coverage, we combine MEV data from three different sources (peer-reviewed work and specialized companies). 

We wrote a crawler to scrape the MEV data from EigenPhi's website and leveraged Etherscan's ``EigenPhi'' label \cite{etherscan_labels} to identify which blocks to scrape from EigenPhi's website. For ZeroMev, we were able to obtain MEV data via their public API.  We slightly modified the scripts of Weintraub et al.~\cite{weintraub2022flash} to be compatible with our post-merge Ethereum client. The scripts detect MEV by analyzing the logs that are triggered by events defined within the smart contracts of the individual platforms (i.e., Uniswap, Aave, Maker, etc.). We collected the logs for transactions in our measurement period from our Ethereum client. 

\T{Sanctioned Transactions.} To identify transactions that involve sanctioned addresses, we first obtain a list of sanctioned addresses from OFAC~\cite{2022OFAC}. We then scan the transaction traces of all transactions in our data collection period to identify any transactions where any nonzero amount of ETH was transferred from or to a sanctioned address. We further scan the logs to identify any transfers of the top five ERC-20 tokens (WETH, USDC, DAI, USDT, and WBTC) from or to any sanctioned address. Additionally, we also monitor all token transfers for TRON as it was sanctioned as of November 2022~\cite{2022ofacnov}. Thus, we are able to obtain lower bounds for transactions that are not OFAC-compliant. Note that we only consider an address sanctioned from the day after it was sanctioned by OFAC, as the OFAC list updates do not have an exact timestamp but are immediately effective~\cite{ofac-compliance}.

\subsection{Ethereum Network}\label{sec:ethereumnetwork}
We obtain Ethereum network data from the Mempool Guru project~\cite{mempool_guru}, which collects all transactions observed in the Ethereum network via their own nodes. In particular, for each transaction included on the Ethereum blockchain, we receive the timestamps at which each of the seven full nodes run by the project first observed the transaction. For the data collection specification, we refer the reader to~\cite{yang2022sok}. This data allows us to distinguish between transactions that were publicly propagated in the network, from those sent through private channels to the creator of the block. 

\subsection{Relay}
\label{sec:relaydata}

Relays act as a bridge between builders and proposers (i.e., validators). Builders submit blocks to relays which are tasked with checking and subsequently forwarding them to proposers. Relays maintain a list of proposers that are currently connected to them. Relays also keep track of which block builders proposed to them and which block the relay chose to be forwarded to the subscribed proposers. Relays expose a public API that builders can leverage to submit blocks and which proposers can use to subscribe to the relay \cite{relay_specs}. This API was proposed by Flashbots~\cite{flashbots}. Table~\ref{tab:relay_list}, provides an overview of the eleven relays that we crawled during our study. These are all of the known relays during the time window of our analysis. We connected to each individual relay and requested for every block number: (1) the final block that was sent to the proposer, (2) a list of blocks that have been submitted by the builders to the relay, and (3) information about proposers that are currently connected to the relay. Most relays are a fork of MEV Boost -- an implementation of PBS built by Flashbots. While Flashbots runs its own relay, anyone can fork Flashbots' implementation and run their own relay. Blocknative is the only relay that uses its own implementation called Dreamboat. MEV Boost and Dreamboat follow the same API, namely Flashbots' relay API specification \cite{relay_specs}.
    
    \begin{table}
        \centering
        \begin{adjustbox}{width=\columnwidth}
        \begin{tabular}{l l c}
                \toprule
                \textbf{Relay Name} & \textbf{Endpoint} & \textbf{Fork} \\
                \midrule
                Aestus & \url{https://aestus.live} & MEV Boost \\
                Blocknative & \url{https://builder-relay-mainnet.blocknative.com} & Dreamboat \\
                bloXroute (Ethical) & \url{https://bloxroute.ethical.blxrbdn.com} & MEV Boost \\
                bloXroute (MaxProfit) & \url{https://bloxroute.max-profit.blxrbdn.com} & MEV Boost \\
                bloXroute (Regulated) & \url{https://bloxroute.regulated.blxrbdn.com} & MEV Boost \\
                Eden & \url{https://relay.edennetwork.io} & MEV Boost \\
                Flashbots & \url{https://boost-relay.flashbots.net} & MEV Boost \\
                GnosisDAO & \url{https://agnostic-relay.net} & MEV Boost \\
                Manifold & \url{https://mainnet-relay.securerpc.com} & MEV Boost \\
                Relayooor & \url{https://relayooor.wtf} & MEV Boost \\
                UltraSound & \url{https://relay.ultrasound.money} & MEV Boost \\
                \bottomrule
        \end{tabular}
        \end{adjustbox}\vspace{2mm}
        \caption{List of PBS relays crawled in this study.}
        \label{tab:relay_list}\vspace{-4pt} 
    \end{table}

\section{Proposer-Builder Separation Landscape}
\label{sec:overview}

\begin{figure}[t]
        \centering
        \includegraphics[width=0.45\textwidth]{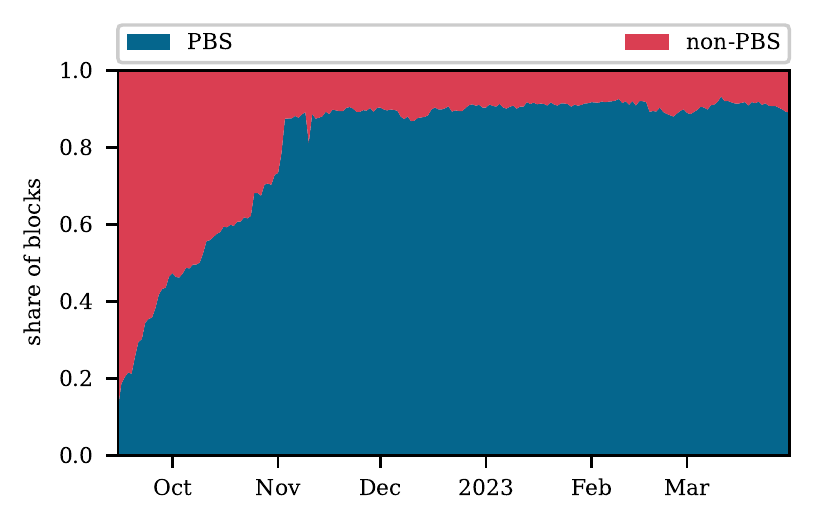}\vspace{-6pt}
        \caption{Daily share of blocks that are built with PBS on Ethereum since the merge. The blue area shows the share of PBS blocks and the red area the share of non-PBS blocks.}
        \label{fig:proportionofMEVBoost}\vspace{-4pt} 
    \end{figure}

We commence the analysis by providing an overview of the adoption of PBS since the merge. Figure~\ref{fig:proportionofMEVBoost} visualizes the proportion of blocks built with PBS over time. We note here that we consider a block to be built through PBS, if it is reported by one of the eleven relays we crawl or if we detect a payment from the builder to the proposer in accordance with the PBS convention. Of all PBS blocks we identify, 99.6\% are claimed by at least one of the relays we crawl, while 92\% exhibit the payment from the builder to the proposer as specified by PBS. Note here that 99.6\% of PBS blocks that do not include the payment from the builder to the proposer have the same builder and proposer address in the PBS data. 

The daily share of PBS blocks rises from around 20\% on 15 September 2022, the day of the merge, to over 85\% on 3 November 2022 (cf. Figure~\ref{fig:proportionofMEVBoost}). From then on out, the daily share of PBS blocks remains relatively stable and ranges between 85\% and 94\% with one exception on 10 November 2022, i.e., the sharp dip in Figure~\ref{fig:proportionofMEVBoost}. This drop in the share of PBS blocks can be traced back to blocks from a builder with incorrect timestamp values being submitted to proposers. Their nodes subsequently rejected the proposed blocks, and the proposers had to fall back to local block production~\cite{nov102022}. 

\subsection{Relays}

We continue by analyzing the relay landscape. First, we provide an overview of the characteristics of the eleven relays we crawled in Table~\ref{tab:relay_overview}. For one, relays differentiate themselves by how they connect to builders. Two relays, namely Blocknative and Eden, do not connect to any external builders, i.e., the relays only forward blocks from their own builders. There are three relays from bloXroute (bloXroute (E), bloXroute (M), and bloXroute (R)) that run their own builders but also connect to external builders. The process of becoming a builder for these relays is not permissionless though. Finally, for the remaining relays, builders can join in a permissionless manner. Further, of these relays, the Flashbots relay is the only one that also runs its own builder. Regarding censorship, several relays announced that they would comply with OFAC sanctions (Blocknative, bloXroute (R), Eden, and Flashbots). Finally, the bloXroute (E) claims to filter out generalized front-running and sandwich attacks. Depending on these announcements by relays as well as possibly additional consideration, validators choose to connect to any number of relays. Recall that as PBS is an opt-in protocol, they do not need to connect to any relay.

    \begin{table}
        \centering
        \begin{adjustbox}{width=\columnwidth}
        \begin{tabular}{@{}l c c c@{}}
                \toprule
                \textbf{Relay Name} & \textbf{Builders}& \textbf{Censorship} & \textbf{MEV Filter} \\
                \midrule
                Aestus & permissionless &  x & x \\
                Blocknative & internal & OFAC-compliant & x \\
                bloXroute (E) & internal \& external & x & front-running\\
                bloXroute (M) & internal \& external & x & x \\
                bloXroute (R) & internal \& external & OFAC-compliant & x\\
                Eden & internal & OFAC-compliant &  x\\
                Flashbots & internal \& permissionless & OFAC-compliant &  x\\
                GnosisDAO & permissionless & x &x \\
                Manifold & permissionless & x &  x\\
                Relayooor & permissionless & x  & x \\
                UltraSound & permissionless & x & x\\
                \bottomrule
        \end{tabular}
        \end{adjustbox}
        \vspace{2mm}
        \caption{Relay overview regarding how relays connect to builders and what their censorship and MEV filtering policies are. We obtain this information directly from the relay websites (cf. Table~\ref{tab:relay_list}).}
        \label{tab:relay_overview}\vspace{-4pt}         
    \end{table}

\begin{figure}[t]
        \centering
        \includegraphics[width=0.45\textwidth]{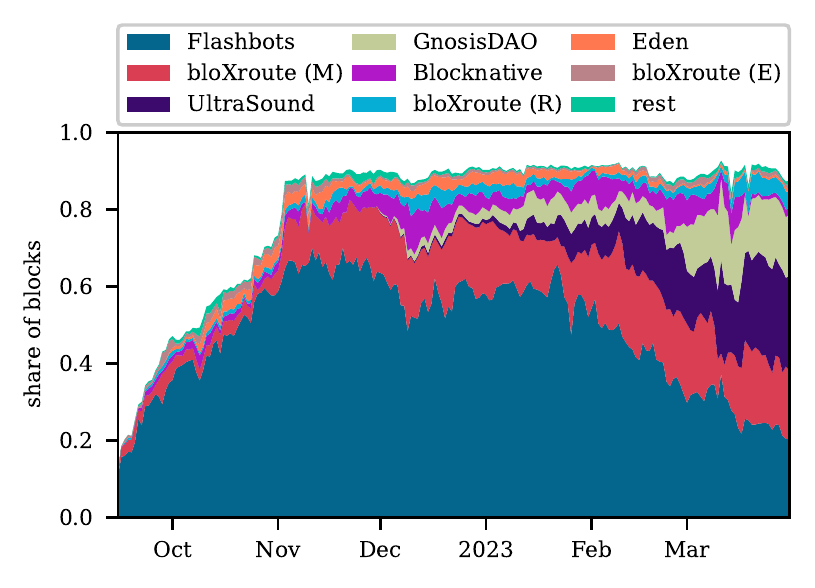}\vspace{-8pt}  
        \caption{Daily share of blocks by each relay over time. We note that in case more than one relay proposes the same block, we attribute the block to each relay equally.}
        \label{fig:relays}\vspace{-6pt}  
    \end{figure}

Figure~\ref{fig:relays} plots the daily share of blocks by each relay during our data collection period. We show the daily block share per relay for the top eight relays (in terms of the number of blocks proposed) and aggregate the daily share of blocks for the remaining relays. We further note that if the same block was proposed by multiple relays, we attribute the block equally to each relay. Around 5\% of all PBS blocks were proposed by more than one relay. Unsurprisingly, the Flashbots relay, i.e., the relay that spearheaded PBS, is the largest. They consistently account for more than half of all blocks proposed through PBS from November 2022 onwards and even account for more than half of all (including non-PBS) blocks between November 2022 and January 2023. Since then, the market share of the Flashbots relay has been decreasing, dropping to 23\% by the end of March. BloXroute (M) is the second largest relay in terms of total number of blocks built and accounts for 20\% of all blocks. All remaining relays are comparatively small initially, but two relays (Ultrasound and GnosisDAO) have experienced a significant increase in their market share since the start of 2023. To summarize, we observe that by the end of March 2023, there are multiple relays with a significant market share while there was only one dominant relay in September 2022. 

\begin{figure}[t]
        \centering
        \includegraphics[width=0.45\textwidth]{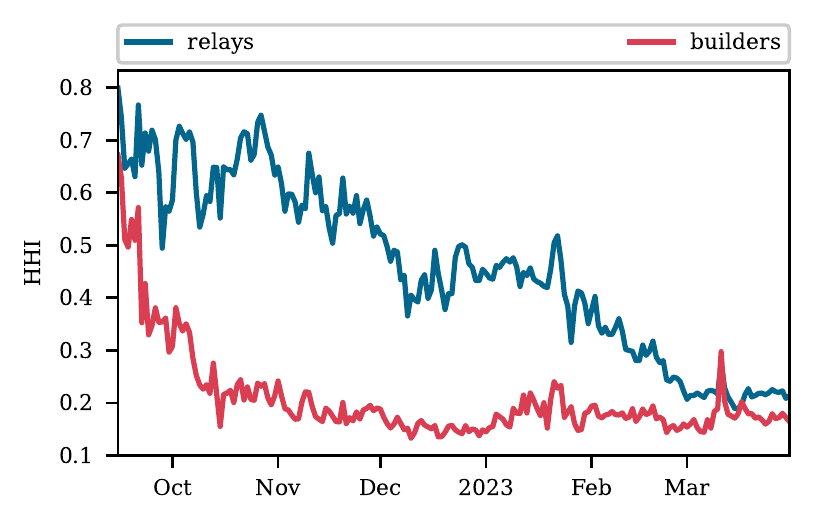}\vspace{-8pt} 
        \caption{Relay and builder HHI over time.}
        \label{fig:HHI}\vspace{-6pt} 
    \end{figure}

To further investigate the decentralization of the relay landscape, we plot the \textit{Herfindahl-Hirschman Index (HHI)} for the relays in Figure~\ref{fig:HHI}. In economics, the HHI is a statistical measure of the concentration of an industry~\cite{rhoades1993herfindahl} and computed as follows
$$\text{HHI} = \sum _{i =1}^{n} (MS_i)^2 \in [0,1],$$
where $n$ is the number of players in a market and $MS_i \in [0,1]$ is the market share of player $i$. The higher the HHI, the higher the concentration of the industry. We note that as opposed to the Gini coefficient, the HHI also takes the number of players into account. In Figure~\ref{fig:HHI}, the blue line shows the concentration amongst the eleven relays. We observe a general downward trend of the HHI with a significant oscillation. The maximum value of the relay HHI is 0.80, and the minimum is 0.19. Generally, an industry with an HHI above 0.15 is said to have a moderate to high concentration. Thus, we summarise that even though the concentration of the relay industry decreases with time, it remains concentrated, i.e., centralized.

\begin{figure}[t]
        \centering
        \includegraphics[width=0.45\textwidth]{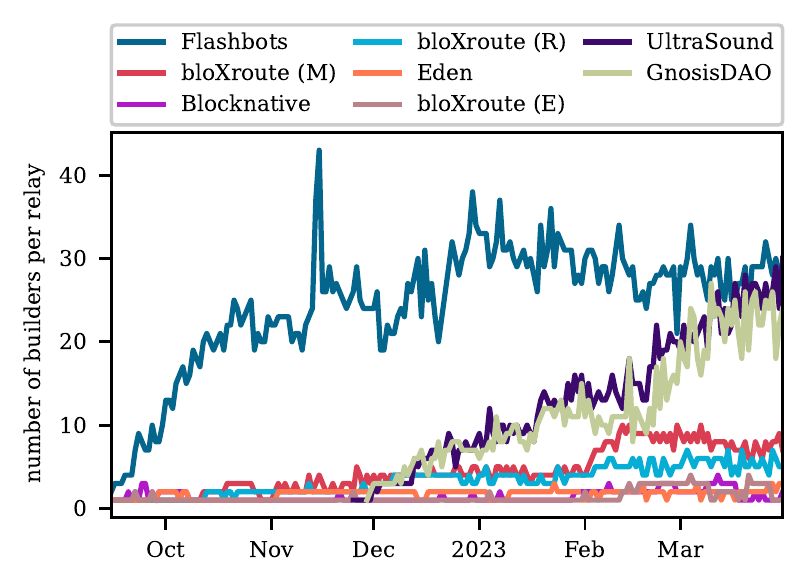}\vspace{-8pt} 
        \caption{Number of builders per relay over time for the top eight relays.}
        \label{fig:builderPerRelay}\vspace{-6pt} 
    \end{figure}

We conclude the general analysis of the relay landscape by analyzing the number of builders per relay. We plot the daily number of different builders sending blocks to each relay over time in Figure~\ref{fig:builderPerRelay}. For the Flashbots relay, the number of builders experienced an initial increase and remained relatively stable from December 2023 onwards at around 30 builders. The number of builders for the second largest relay (bloXroute (M)) was initially very low but experienced a sharp and consistent increase. Finally, we remark that the number of builders for both the UltraSound and GnosisDAO relays is very similar and rising over time (almost overtaking Flashbots' numbers by the end of our study period), and the number of builders for the remaining relays has stayed comparatively low. We further observe that the number of builders is increasing, especially for relays that are permissionless (cf. Table~\ref{tab:relay_overview}). Additionally, we notice that permissionless relays with a higher number of builders have a more significant market share (cf. Figure~\ref{fig:relays}).

\subsection{Builders}

\begin{figure}[t]
        \centering
        \includegraphics[width=0.45\textwidth]{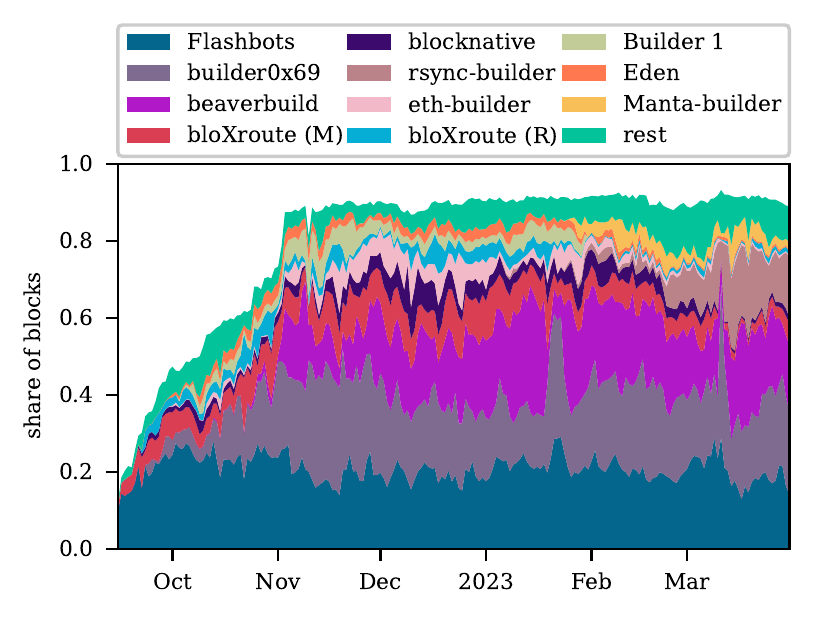}\vspace{-8pt} 
        \caption{Daily share of blocks by each builder over time.}
        \label{fig:builders}\vspace{-6pt} 
    \end{figure}

Previously, we analyzed the number of builders per relay but builders themselves can also be connected to multiple relays at a time. Thus, in this section, we take a look at the builder landscape in detail. Note that throughout we will identify individual builders by their builder public key. We further cluster together builders that use the same fee recipient address. We provide a map from the builder name to the builder fee recipient address(es) and the builder public key(s) in Appendix~\ref{app:builderaddress} in Table~\ref{tab:builderaddress}. Figure~\ref{fig:builders} visualizes the daily share of blocks by each of the biggest eleven builders, in terms of the total number of blocks. We notice that the top three builders, i.e., Flashbots, builder0x69, and beaverbuild, consistently account for more than half of all blocks together from November 2022 onwards. The remaining builders all individually account for less than 15\% of blocks on each day. Thus, the builder landscape, just as the relay landscape, appears to be dominated by a small number of large players. 

This picture cements itself when we calculate the daily builder HHI (cf. red line in Figure~\ref{fig:HHI}). In total, there were 133 unique builders, and the builder HHI oscillates between 0.13 and 0.67 with a mean of 0.21. We further observe a very significant downward trend in the concentration initially. From November onwards, the concentration of the builder landscape remains relativity stable at around 0.17. We note that HHI between 0.15 and 0.25 indicates moderate concentration, while only HHI values below 0.15 indicate an unconcentrated industry. Thus, we conclude that the builder landscape is also quite concentrated but generally less so than the relay industry. 

\section{PBS Impact on Block Composition}
\label{sec:blockcomp}

In the following, we investigate the block composition of PBS blocks by individual builders and in comparison to non-PBS blocks. We place a particular focus on the block value to analyze whether PBS ``decentralizes transaction validation by not giving large entities an advantage in block building'' as indented by its design goals~\cite{2022pbs}. The idea is that if validators had to build blocks themselves, large entities would be at an advantage. As a result of the high complexity of block building, large entities would be expected to build comparatively more profitable blocks and would grow even bigger --- increasing transaction validation centralization. 

\subsection{Block Value}
\begin{figure}[t]
        \centering
        \includegraphics[scale =1]{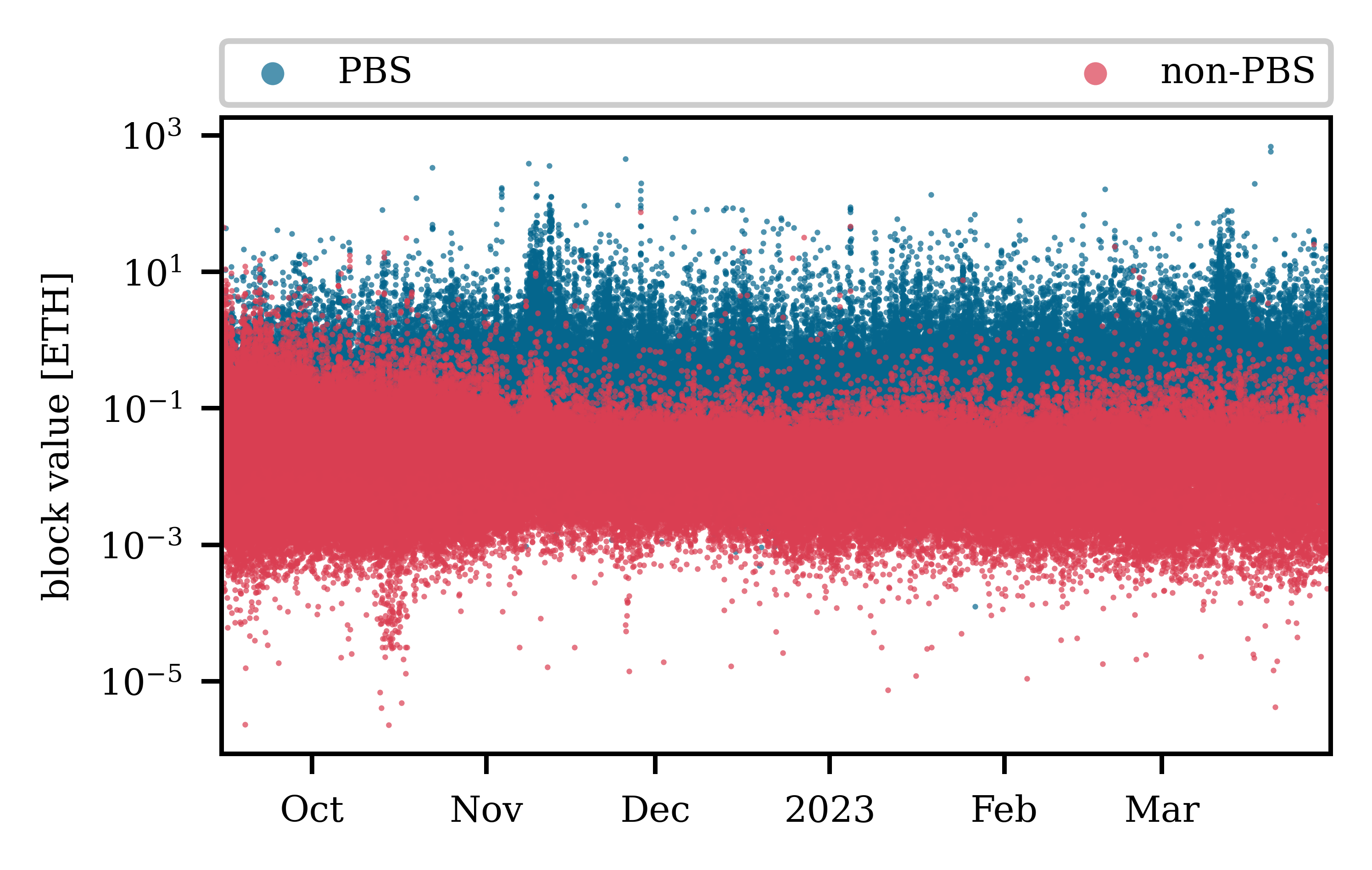}\vspace{-8pt} 
        \caption{Block value, i.e., priority fees and direct transfers, across time. Each dot represents a block. PBS blocks are shown in blue and non-PBS blocks are shown in red. Notice that the value of PBS blocks tends to be higher than that of non-PBS blocks.}
        \label{fig:blockprofit}\vspace{-6pt} 
    \end{figure}

Recall that we define block value as the amount of user-generated reward available in a block (i.e., priority fees and direct transfers) and shared by the builder and proposer. We compare the block value of PBS and non-PBS blocks in Figure~\ref{fig:blockprofit}. Notice that the block value for PBS blocks is consistently significantly higher than for non-PBS blocks. The gap appears to grow with time which might be due to a higher level of PBS adoption or increasing sophistication of PBS  block builders. Regardless, it appears that professionalized builders are at an advantage concerning value extraction compared to proposers who are building their blocks outside of PBS.

\begin{figure}
        \centering
        \includegraphics[scale =1]{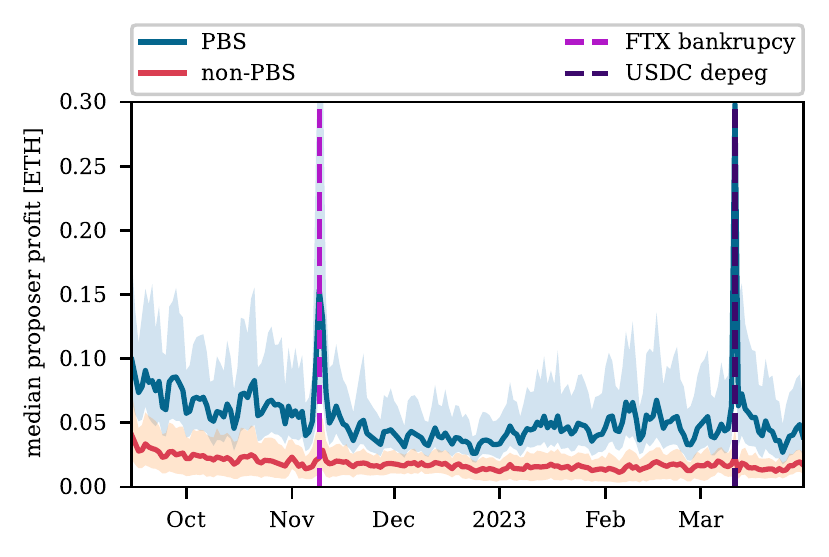}\vspace{-8pt} 
        \caption{Daily median proposer profit for PBS blocks (blue line) and non-PBS blocks (red line). The shaded area around the lines indicates the 25th to 75th percentile of proposer profits.}
        \label{fig:medianproposerprofit}\vspace{-6pt} 
    \end{figure}

Additionally, we compare the daily median proposer profits (amount the builder pays out to the proposer) for proposers that use PBS and those that do not use PBS (total block value) in Figure~\ref{fig:medianproposerprofit}. We find that PBS proposers have significantly higher median profits (blue line) than non-PBS proposers (red line). This difference is largest on high MEV days, i.e., FTX bankruptcy (violet dashed line) and USDC depeg (purple dashed line).  Even more startlingly, the 25th percentile of PBS proposer profits (bottom of blue area) is generally above the 75th percentile of non-PBS proposers (top of red area). Thus, it appears that proposers will generally make greater profits if they use PBS as opposed to building their own blocks --- an indication that PBS achieves its goal of not giving institutional validators an advantage over hobbyist validators in value extraction. 

\subsection{Distribution of Value}

\begin{figure}
        \centering
        \includegraphics[scale =1]{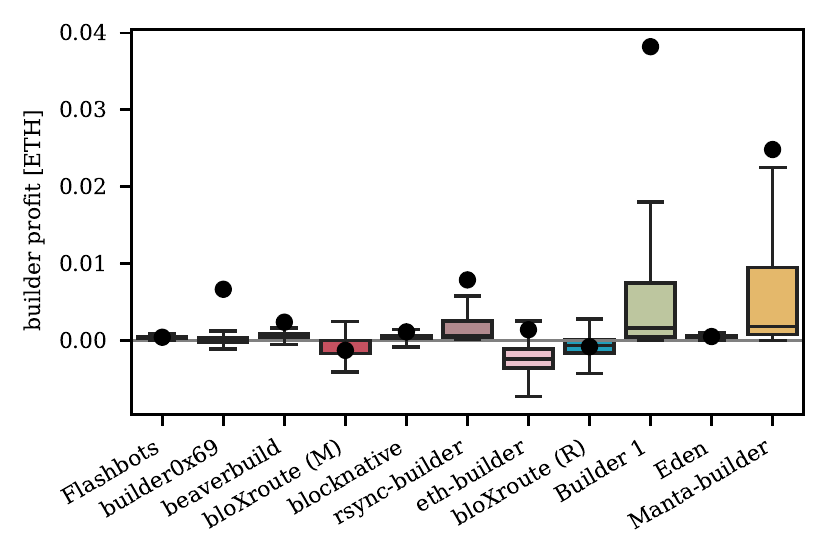}\vspace{-8pt} 
        \caption{Box plot of builder profits. We sort builders by their size and indicate the mean with a black dot.}
        \label{fig:builderprofit}\vspace{-6pt} 
    \end{figure}

    In the following, we analyze the distribution of a block's value between block builders and proposers. Figure~\ref{fig:builderprofit} visualizes the builder profit (i.e., priority fees and direct transfers subtracted by the builder payment to the proposer) as a box plot for each of the top eleven builders. One immediately notices that the builder profit varies significantly between builders. The Flashbots, Eden, and blocknative builders appear to follow a similar strategy. The variance of their profit per block is very small, and the mean profit, indicated by the black dot, is between 0.0004 and 0.001 ETH for each of them. We also observe a similar pattern for Builder 1, rsync-builder, and Manta-builder. These three builders are the most profitable builders, with more than 0.0075 ETH average profit per block each. Further, they, along with the previously mentioned builders, do not tend to subsidize (i.e., make a negative profit) on blocks.

    For the remaining five builders, the profit per block is regularly negative, i.e., they are subsidizing the blocks. However, while the builder0x69, beaverbuild, and eth-builder frequently subsidize blocks, their mean profit is positive. Thus, while subsidizing some blocks, potentially to encourage transaction flow from searchers and build trust with them, they make up for it on high-value blocks. The mean profit of two bloXroute builders, on the other hand, is not positive. We conclude that they either subsidize blocks or are making a profit in ways that we cannot detect. The bloXroute founder has commented on this phenomenon and noted that ``the bloXroute builder can sustainably subsidize their blocks using earnings from other business revenue streams''~\cite{bloxrouteceo}. He, however, did not further hint at the origin of these revenue streams.

\begin{figure}
        \centering
        \includegraphics[scale =1]{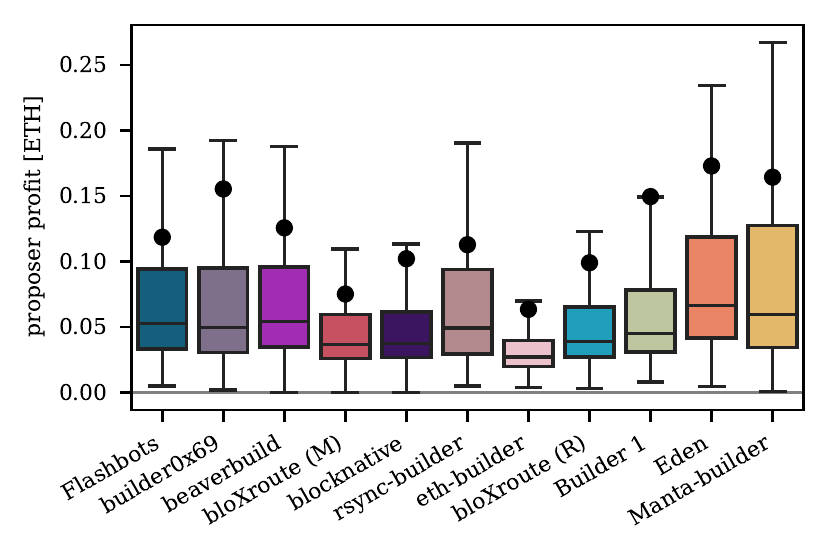}\vspace{-8pt} 
        \caption{Box plot of proposer profits for each builder. We sort builders by their size and indicate the mean with a black dot.}
        \label{fig:proposerprofitprofit}\vspace{-6pt} 
    \end{figure}

    When turning to the proposer profits (cf. Figure~\ref{fig:proposerprofitprofit}), the picture looks more uniform. In general, each builder is offering the proposers similar profits for their blocks. We note that we observe differences of more than a factor of two in mean proposer profits for different builders. However, we believe that, in part, these differences likely stem from the builders being active at different times. We also note the proposer profits are highly skewed, i.e., the mean is significantly higher than the median, which is due to the existence of large MEV opportunities that come about rarely and drive up the mean. Combining our builder profit and proposer profit analysis, we conclude that the proposers' profits are more than a factor of ten higher on average than the builder profits. We provide an additional analysis of this in Appendix~\ref{app:profit}.

    Finally, we analyze what proportion of the value promised by relays to the proposers is actually delivered. Recall that proposers initially only receive a blinded block along with the value of the block, i.e., the profit that is promised to them from the relays. Thus, when choosing to sign a block, they pick the one with the highest value and trust the relay that this value will be delivered. The left side of Table~\ref{tab:ofac} shows, for the blocks that were proposed, the total value delivered to the proposers along with the total value that was promised. We compare the two values and calculate the share of the promised value that was delivered in the third column of Table~\ref{tab:ofac}. Astonishingly, each relay, with the notable exception being Aestus, has not delivered the full promised value and thus broke the trust they enjoyed from the proposers.
    
    Despite this, we note that all but two relays delivered more than 99.8\% of the value. The two exceptions are the Eden relay, which only delivered 93.8\% of the promised value, and the Manifold relay, which only delivered 19.9\% of the promised value. For the Eden relay, the vast majority of the missing value stems from a single block. The relay announced the value of block 15,703,347 to be 278.29 ETH but only delivered 0.16 to the proposer. We note that the Eden relay did not deliver the full value in 0.05\% of blocks (cf. fourth column of Table~\ref{tab:ofac}). Similarly, for Manifold, a large chunk of the missing value stems from the same day --- 15 October 2022. On this day, a builder noticed that the Manifold relay was not checking the block rewards of the blocks that it was receiving~\cite{Postmortemmanifold}. The builder, thus, submitted blocks with wrongly declared rewards, and 184 of these blocks made it onto the blockchain. Then, the profit from these blocks went to the builder, and the proposers were left with nothing. Across all relays, 98.7\% of the promised value arrived at the proposers and 0.86\% of blocks did not deliver their promised value.

\begin{table*}[!htb]
\scriptsize

      \centering
      \begin{adjustbox}{width=\linewidth}
        \begin{tabular}{@{}l | rrrr|rr@{}}
\toprule
{} &  delivered value [ETH] &  promised value [ETH] &       share of value [\%] &   share over-promised of blocks [\%] & sanctioned blocks& share of sanctioned blocks [\%]  \\
\midrule
Aestus        &       404.582057 &      404.582057 &      100.000000 &        0.030921 &      35 &   1.082251 \\
\textit{Blocknative}   &      6086.147269 &     6087.225188 &       99.982292 &        3.553043 &    1001 &   1.807610 \\
bloXroute (E) &      1027.721708 &     1028.849142 &       99.890418 &        4.449079 &     780 &   5.420431 \\
bloXroute (M) &     19248.859647 &    19250.965641 &       99.989060 &        2.723592 &   10038 &   5.375475 \\
\textit{bloXroute (R)} &      3689.983494 &     3690.389671 &       99.988994 &        0.113681 &     254 &   0.824756 \\
\textit{Eden}          &      4204.908110 &     4483.541341 &       93.785421 &        0.047946 &      81 &   0.323586 \\
\textit{Flashbots}     &     90812.869179 &    90818.879444 &       99.993382 &        0.032546 &    1451 &   0.210801 \\
GnosisDAO     &     12461.847484 &    12462.572553 &       99.994182 &        0.894293 &    2225 &   2.956300 \\
Manifold      &       416.283800 &     2095.806623 &       19.862701 &        6.880409 &    1012 &  14.356646 \\
Relayooor     &       319.504570 &      319.608181 &       99.967582 &        2.095704 &     162 &   5.658400 \\
UltraSound    &     13970.478128 &    13971.955037 &       99.989429 &        0.953061 &    3289 &   3.309419 \\
\midrule
PBS           &    152643.185446 &   154614.374878 &       98.725093 &        0.855103 &   20328 &   1.710460\\
\bottomrule
\end{tabular}
  \end{adjustbox}\vspace{2mm}
    \caption{The left side of the table compares the block value delivered to proposers with the promised value for each relay. The right part side of the table shows the number and percentage of blocks built by the relay that include OFAC non-compliant transactions. Relays that announce to be OFAC-compliant are highlighted in italics.}
    \label{tab:ofac}\vspace{-8pt} 
\end{table*}

\subsection{Block Contents}
\begin{figure}
        \centering
        \includegraphics[scale =1]{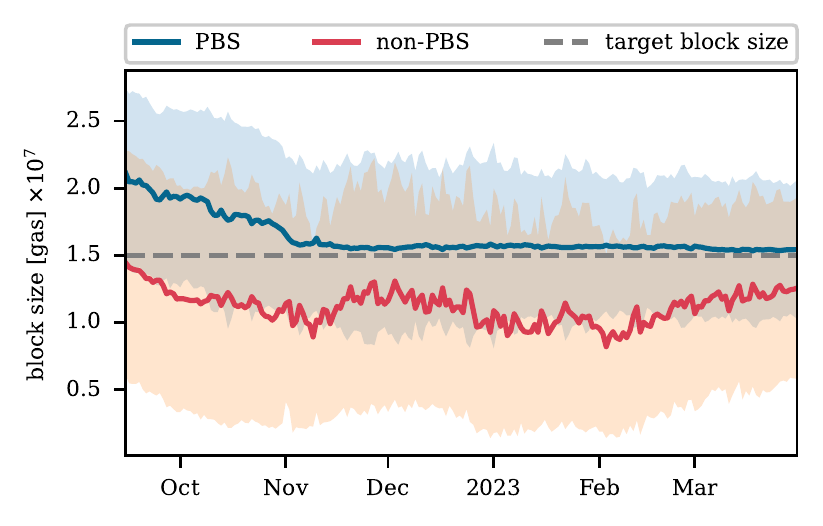}\vspace{-8pt} 
        \caption{Mean daily block size for PBS (shown in blue) and non-PBS (shown in red) blocks along with the standard deviation. We indicate Ethereum's target block size by the dashed line.}
        \label{fig:gasused}\vspace{-6pt} 
    \end{figure}

    We further analyze how blocks are composed by looking at how builders are including transactions in blocks. In Figure~\ref{fig:gasused} we look at the daily average block size, i.e., the gas usage, of PBS blocks vs. non-PBS blocks. We see that PBS blocks start off higher than Ethereum's target block size with $2\cdot 10^7$ gas right after the merge and then drop with time until mid-November (blue line in Figure~\ref{fig:gasused}). From then on the mean daily block size of PBS block is stable and hovers slightly above Ethereum's target block size of $1.5\cdot 10^7$ gas. However, even though the daily average is stable, the standard deviation (the blue-shaded area) is significant. For non-PBS blocks the mean block size, the red line in Figure~\ref{fig:gasused}, is continuously below the target size and exhibits greater fluctuations than that of PBS blocks. Similarly, the standard deviation of non-PBS blocks is slightly larger than that of PBS blocks. Thus, PBS blocks tend to, at least on average, be fuller and have a more consistent size than non-PBS blocks.

\begin{figure}
        \centering
        \includegraphics[scale =1]{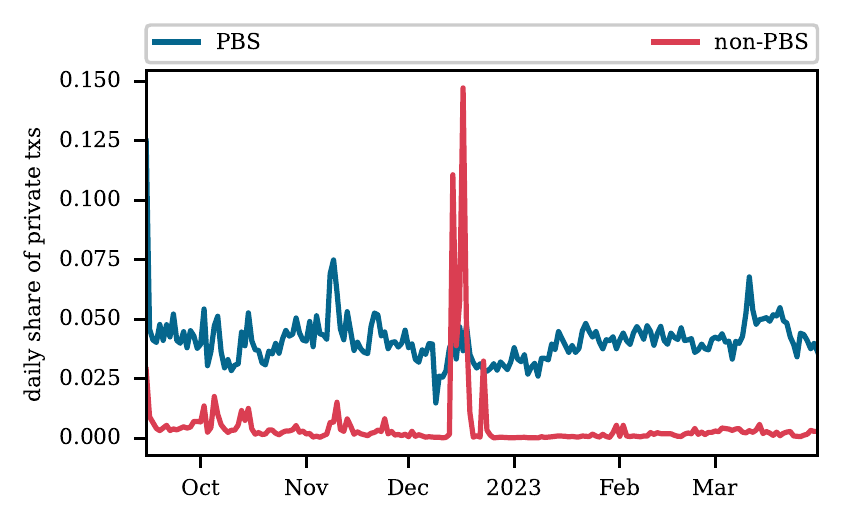}\vspace{-8pt} 
        \caption{Daily share of private transactions included in PBS and non-PBS blocks.}
        \label{fig:percentPrivate}\vspace{-6pt} 
    \end{figure}

    We also explore the fraction of transactions in blocks that are coming from the P2P network (i.e., observed in the public mempools), versus sent via private avenues. In Figure~\ref{fig:percentPrivate} we observe that private transactions are largely being included by PBS block builders. These are likely from the searchers of builders themselves, from searchers sending transactions directly to builders  and privacy services to prevent attacks  (e.g., Flashbots~\cite{flashbots} and Bloxroute\cite{klarman2018bloxroute}). 

    Note that the peak in private transaction share of non-PBS blocks in December is almost entirely made up of a single sender ($s$) and receiver  ($r$) pair, where
    \begin{itemize}
        \item $s$: \texttt{0x4d9ff50ef4da947364bb9650892b2554e7be5e2b}
        \item $r$: \texttt{0x0b95993a39a363d99280ac950f5e4536ab5c5566}.
    \end{itemize}
    Both addresses are from Binance and the transactions are simple ETH transfers. In total, these private transactions from Binance make up for more than 88\% of all private transactions in non-PBS blocks in December 2022. Of these, more than 90\% are 75 blocks proposed by AnkrPool proposers. Thus, it appears that AnkrPool proposers were receiving private transactions from Binance for around two weeks in December 2022. 

\subsection{MEV value}

    To conclude our exploration of the impact of PBS on block composition, we take an in-depth look at the number of and the profit from the MEV transactions in PBS and non-PBS blocks. Note that throughout, we will focus on the three most well-known and frequent types of MEV: sandwich attacks, cyclic arbitrage, and liquidations.

    We start by commenting on the reliability of the MEV filtering policy by the bloXroute (E) relay. Recall from Table~\ref{tab:relay_overview} that the bloXroute (E) claims to filter out front-running transactions --- including both generalized front-running and sandwich attacks. While it is non-trivial and not always possible to detect generalized front-running as the front-run transactions will likely not be included on the blockchain, we can use our MEV data to identify how many sandwich attacks were in blocks from the bloXroute (E) relay. We find that since the merge, the bloXroute (E) has included 2,002 sandwich attacks and therefore conclude that the filtering performed by the relay exhibits significant gaps.

\begin{figure}
        \centering
        \includegraphics[scale =1]{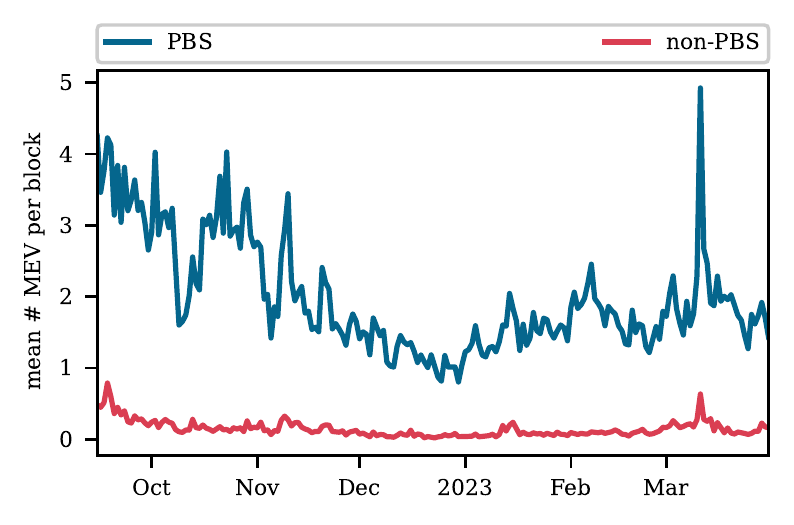}\vspace{-8pt} 
        \caption{Mean number of MEV transactions per block for PBS (blue line) and non-PBS (red line) blocks.}
        \label{fig:mevtxs}\vspace{-6pt} 
    \end{figure}

    We continue with a more general analysis of MEV in PBS and non-PBS blocks. Figure~\ref{fig:mevtxs} shows the mean number of MEV transactions per block for PBS blocks (blue line) and non-PBS blocks (red line). One notices that the number of MEV transactions is significantly higher in PBS blocks than in non-PBS blocks. Thus, it appears that builders have better connections to searchers and their in-flow of MEV transactions than proposers building their own blocks. In Appendix~\ref{app:mev}, we further break up the MEV transactions into sandwich attacks, cyclic arbitrage, and liquidations to compare the prevalence of each type of MEV in PBS and non-PBS blocks.

\begin{figure}
        \centering
        \includegraphics[scale =1]{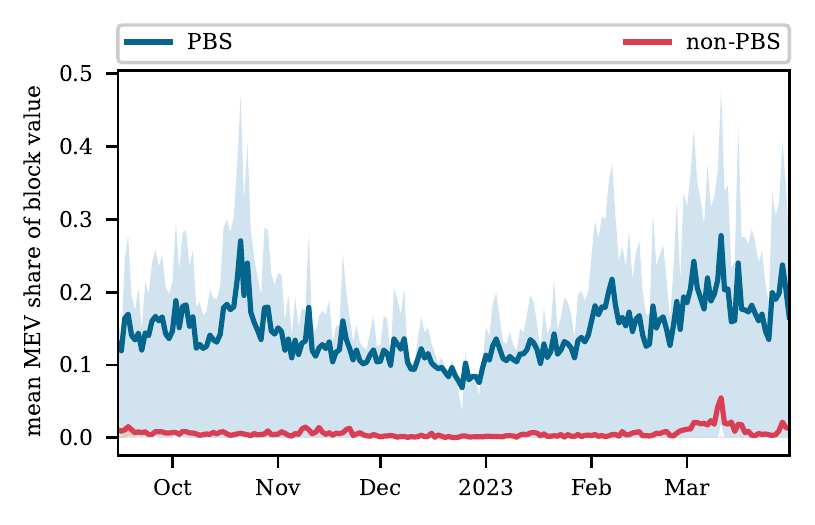}\vspace{-8pt} 
        \caption{Daily mean of block value that stems from MEV for PBS blocks (blue line) and non-PBS blocks (red line). The shaded area around the lines indicates the 25th to 75th percentile of block value that comes from MEV.}
        \label{fig:mev}\vspace{-6pt} 
    \end{figure}

    Generally, MEV transaction flow is regarded as crucial for both builders and proposers as it accounts for a significant proportion of block value. We plot the daily mean share of block value that comes from MEV transactions for PBS and non-PBS blocks in Figure~\ref{fig:mev}. Note that the shaded area indicates the 25th and 75th percentile. Similarly to our previous analysis, we find that while MEV makes up a significant share of the block value for PBS blocks, 14.4\% on average, very little of the block value can be attributed to MEV for non-PBS blocks.

    \section{Censorship Resistance}
\label{sec:censorship}

\begin{figure}
        \centering
        \includegraphics[width=0.45\textwidth]{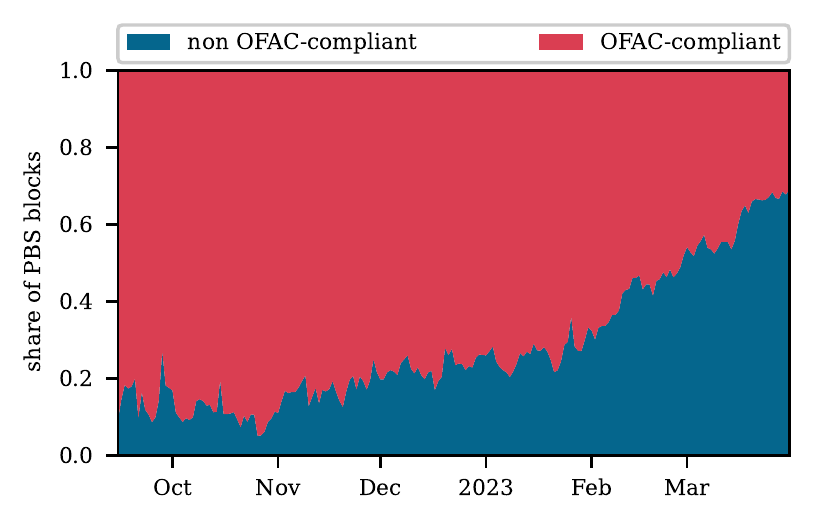}\vspace{-8pt} 
        \caption{Share of PBS blocks that are produced by non-OFAC-compliant versus OFAC-compliant relays.}
        \label{fig:relaysofac}\vspace{-6pt} 
    \end{figure}

We conclude the analysis by investigating whether PBS achieves its second design goal: preventing censorship. The idea is that since the block proposer only receives the block header and is not aware of the block contents when signing the block, it is not possible for powerful organizations to pressure proposers into complying with their sanctions~\cite{2022pbs}. Throughout this section, we focus on OFAC sanctions when discussing censorship, given that it is the most prominent case study of censorship in the Ethereum ecosystem. As we already pointed out (cf. Table~\ref{tab:relay_overview}), a number of relays announce that they are OFAC-compliant, i.e., they censor transactions in accordance with OFAC sanctions. Thus, we start by analyzing the daily share of blocks produced by relays that self-report themselves as censoring in Figure~\ref{fig:relaysofac} to obtain an overview of the proportion of blocks that stem from censoring relays. Notice that initially, the share of PBS blocks built by OFAC-compliant relays is more than 80\%. From November on, this share starts to decrease and reaches just north of 45\% by the end of March. However, even though the proportion of PBS blocks built by non-censoring relays increases, a significant proportion of PBS blocks are built by censoring relays.

\begin{figure}
        \centering
        \includegraphics[width=0.45\textwidth]{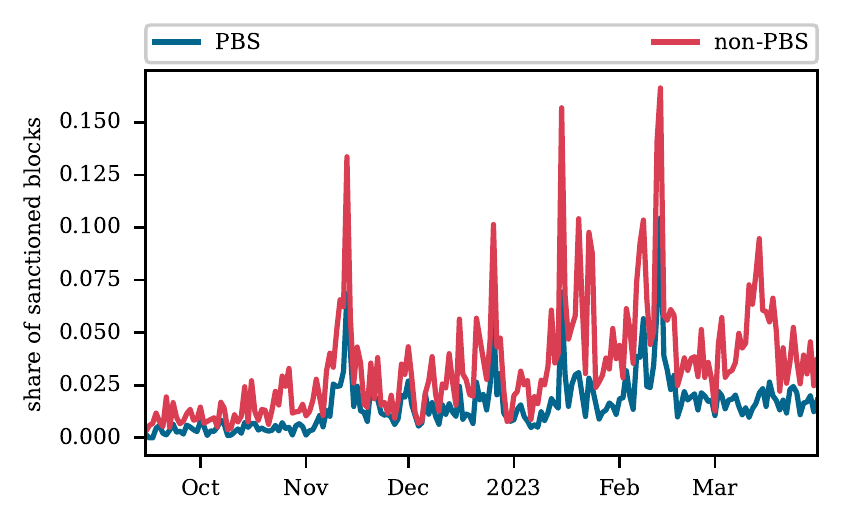}\vspace{-8pt} 
        \caption{Daily share of PBS and non-PBS blocks that include transactions that do not comply with OFAC sanctions.}
        \label{fig:percentOfac}\vspace{-6pt} 
    \end{figure}

To further investigate whether PBS helps prevent censorship, we analyze the share of blocks that include transactions that are not OFAC-compliant in Figure~\ref{fig:percentOfac}. In particular, we compare PBS (shown in blue) and non-PBS (shown in red) blocks and observe that in general, a more significant proportion of non-PBS blocks include transactions that are not OFAC-compliant in comparison to PBS blocks. Thus, it does not appear that PBS is preventing censorship as it was envisioned but instead is linked to more censored blocks. Additionally, PBS gives an avenue for proposers who wish to comply with OFAC sanctions to only connect to relays that promise to comply with OFAC sanctions. Thus, we believe that there are incentives for at least some relays to comply with OFAC sanctions.

Finally, we point out that even though some relays advertise themselves to be OFAC-compliant, they do not always uphold their promises. On the right side of Table~\ref{tab:ofac}, we show the number of blocks along with the share of total blocks that each of the biggest eight relays included, which contain transactions that are not OFAC-compliant. We observe that every single relay, regardless of whether they promised to adhere to OFAC sanctions or not, has included transactions that are not OFAC-compliant. The relays that self-report being OFAC-compliant are highlighted in italics in Table~\ref{tab:ofac}. Further, we find the most significant gaps in their filtering to follow updates of the OFAC sanctions list. For instance, new Ethereum addresses were added to the OFAC sanctions list on 8 November 2022~\cite{2022ofacnov}, but the OFAC blacklist of the Flashbots relay was only updated on 10 November 2022~\cite{2023ofacblacklist}. Further, updates to the OFAC list on 1 February 2023~\cite{2023ofacfeb} were still not reflected in the OFAC blacklist of the Flashbots relay on 1 May 2023.

Though we find that all relays that promise to be OFAC-compliant let through non-OFAC-compliant transactions, the share of blocks that are not OFAC-compliant from these relays is significantly smaller than from the other relays. For instance, while more than 14\% of blocks from Manifold do not comply with OFAC sanctions, the same only holds for 0.21\% of blocks from the Flashbots relay. Thus these relays are largely reliable in their filtering, modulo days when there are updates to the OFAC sanctions list.

\section{Related Work}
\label{sec:relatedwork}
\T{Mining Pool Distribution.} The value proposition of cryptocurrencies is to facilitate payments in a decentralized manner. Gencer et al.~\cite{gencer2018decentralization} were the first to study the level of decentralization of Bitcoin and Ethereum miners. Their work shows that the mining process of both blockchains is fairly centralized. Subsequent studies of the mining power distribution on Ethereum by Kiffer et al.~\cite{kiffer2021under} and Lin et al.~\cite{lin2021measuring} support these findings. Our work studies Ethereum post-merge and finds that block building via PBS continues to exhibit significant levels of centralization. 

\T{Transaction Inclusion.} With limited block space, users traditionally bid for inclusion in blocks via the transaction fee mechanism. Previous works by Messias et al. \cite{messias2021selfish} and Messias et al. \cite{messiasdissecting}, explore the lack of transparency in fee auctions for Bitcoin and Ethereum. Both works show that often transactions don't follow a fee-based ordering and that the lack of transparency leads to higher user costs and miner profits. Earlier work by Pappalardo et al. also explored the inefficiencies of transaction inclusion times in the Bitcoin network \cite{pappalardo2018blockchain}.

\T{Maximal Extractable Value.} The increasing complexity of block building is in large part a result of MEV. Eskandari et al.~\cite{eskandari2020sok} and Daian et al.~\cite{daian2020flash} offer an early description of MEV. A succeeding stream of literature quantifies the amount of MEV on the Ethereum blockchain~\cite{torres2021frontrunner,qin2022quantifying,wang2022cyclic} and, thereby, demonstrates the immense scale and impact of MEV. 

In their pre-merge exploration of MEV, Piet et al.~\cite{piet2022extracting} place a particular focus on the value distribution between searchers and miners. They find that the majority of MEV profits go directly to the miners. In our exploration of PBS, we analyze the profit share between builders and proposers post-merge and find that most of the profit goes to the proposers. 

Given the losses faced by users as a result of MEV, multiple approaches emerged to prevent or lessen the impact of MEV. We refer the reader to~\cite{heimbach2022sok} for an overview of these approaches. The approach with the most significant level of adoption pre-merge was MEV auction platforms. These platforms aimed to protect the privacy of the user transactions pre-execution, i.e., provide front-running protection, and move the priority gas auction off-chain. Flashbots~\cite{flashbots}, was the biggest MEV auction platform. Weintraub et al.~\cite{weintraub2022flash} provide an empirical analysis of the Flashbots platform. Their work shows that powerful miners doubled their profits with Flashbots, while the profits for searchers decreased. In contrast to their work, we study PBS and the progression of Flashbots, post-merge. Our analysis also indicates that proposers, the new miners, profit by joining PBS. 

\T{Proposer-Builder Separation.} An early exploration of PBS is provided by Yang et al.~\cite{yang2022sok}. Their work places a particular focus on the censorship of sanctioned transactions in the PBS ecosystem. They find that in the first couple months of PBS, sanctioned transactions experienced waiting times that were, on average, 68\% longer than those of regular transactions. Further, we became aware of two simultaneous works~\cite{wahrstatter2023time,wahrstatter2023blockchain} that analyze the PBS landscape and its impact on censorship. Our work provides a more comprehensive analysis of the decentralization of the PBS ecosystem as well as the impact of PBS on block content and censorship.

\section{Concluding Discussion}
\label{sec:discussion}

The full integration of PBS into Ethereum clients is currently on the Ethereum road map~\cite{2022pbs}. Thus, it is essential to analyze the current state of PBS during its opt-in phase to investigate how its realities measure up to its promises.

\T{Decentralization of Transaction Validation.} Our work investigates whether PBS achieves its design goal of decentralizing block validation by not giving large entities an advantage in block building. Through our in-depth analysis of the impact of PBS on the block (value) composition, we come to the conclusion that professionalized builders have a clear advantage in building profitable blocks. Thus, PBS gives validators access to competitive blocks and can prevent ``hobbyists from being out-competed by institutional players that can better optimize the profitability of their block building''~\cite{2022pbs}. 

At the same time, we observe a significant centralization for both the relays and builders. Though the main PBS article on the Ethereum Foundation's website suggests that there is no harm in centralizing block building, as long as the validators are decentralized~\cite{2022pbs}, there are still concerns regarding the potential harm of builder centralization. For one, Vitalik Buterin~\cite{centralizebuilding} commented on the need to prevent new types of censorship vulnerability stemming from builder centralization. Our work suggests this is still an open problem.

\T{Censorship Resistance.} With our work, we also shed light on the extent to which PBS achieves its second design goal: preventing censorship. We find no signs of PBS preventing censorship in practice. In fact, we consistently observe that PBS blocks are less likely to include transactions from sanctioned addresses than non-PBS blocks. Thus, our analysis concludes that PBS currently falls short of its second design goal. 

\T{Trust in Relays.} To conclude, we comment on the trust in relays that is required by the current PBS implementation. Presently, relays are trusted by both builders, to keep their blocks blinded until they are signed, and by proposers, to deliver the promised value and adhere to any additional promises made concerning censorship and MEV filtering, for instance. Our data provides us with the opportunity to analyze whether relays have broken the trust placed on them by proposers. Astonishingly, we find instances of all but one relay falling short of their promises. We observe relays not delivering the full promised value, as well as significant gaps in their censorship and MEV filtering policies. 

The current plan for a native implementation of PBS into the Ethereum protocol reduces the aforementioned trust assumptions by eliminating the need for relays~\cite{pbsproposal}. However, there are still security concerns regarding the present proposal~\cite{pbsproposal2}. The proposal is also restricted to ensuring that the value is delivered but does not address the other aspects. Thus, we believe that we are still far from eliminating the need for relays. Meanwhile, the lack of incentives for relays not to misbehave is an open problem: currently, the revenue of these key players is not part of the PBS design. 

\section{Acknowledgements}
We thank Balakrishnan Chandrasekaran and the anonymous IMC 2023 reviewers for their useful suggestions. L. Kiffer contributed to this project while under an armasuisse Science and Technology CYD Distinguished Postdoctoral Fellowship.
This work was supported by the Zurich Information Security \& Privacy Center (ZISC).
\bibliographystyle{ACM-Reference-Format}
\balance
\bibliography{references}
\clearpage
\nobalance
\section{Ethics}
    Our datasets deal exclusively with publicly available data (i.e., Ethereum blockchain data), MEV and relay data that can be queried by any entity, and mempool data that is logged from the public P2P network (which we use only to mark transactions as private or public). Additionally, all Ethereum addresses presented in this paper are from large entities that purposefully make their addresses publicly known. As such, this paper does not raise any ethical concerns.

\section{Builder Address and Public Key} 
    \label{app:builderaddress}
\newcommand{\specialcell}[2][c]{%
  \begin{tabular}[#1]{@{}c@{}}#2\end{tabular}}
\begin{table*}[b]
\scriptsize

\centering

\begin{adjustbox}{width=\linewidth}

\begin{tabular}{@{}l l l @{}}
\toprule
Name & Address & Public Key  \\
\midrule
beaverbuild & \texttt{0x95222290dd7278aa3ddd389cc1e1d165cc4bafe5} &
\specialcell{\texttt{0x96a59d355b1f65e270b29981dd113625732539e955a1beeecbc471dd0196c4804574ff871d47ed34ff6d921061e9fc27} \\
\texttt{0xb5d883565500910f3f10f0a2e3a031139d972117a3b67da191ff93ba00ba26502d9b65385b5bca5e7c587273e40f2319} \\
\texttt{0x8dde59a0d40b9a77b901fc40bee1116acf643b2b60656ace951a5073fe317f57a086acf1eac7502ea32edcca1a900521}\\
\texttt{0xaec4ec48c2ec03c418c599622980184e926f0de3c9ceab15fc059d617fa0eafe7a0c62126a4657faf596a1b211eec347} } \\

\midrule

bloXroute (M)  &\texttt{0xf2f5c73fa04406b1995e397b55c24ab1f3ea726c} &\specialcell{\texttt{0x94aa4ee318f39b56547a253700917982f4b737a49fc3f99ce08fa715e488e673d88a60f7d2cf9145a05127f17dcb7c67}\\
\texttt{0x976e63c505050e25b70b39238990c78ddf0948685eb8c5687d17ba5089541f37dd3c45999f2db449eac298b1d4856013} \\
\texttt{0x8b8edce58fafe098763e4fabdeb318d347f9238845f22c507e813186ea7d44adecd3028f9288048f9ad3bc7c7c735fba} \\
\texttt{0xaa1488eae4b06a1fff840a2b6db167afc520758dc2c8af0dfb57037954df3431b747e2f900fe8805f05d635e9a29717b} \\

}\\

\midrule

bloXroute (R) & \texttt{0x199d5ed7f45f4ee35960cf22eade2076e95b253f} & \specialcell{\texttt{0x80c7311597316f871363f8395b6a8d056071d90d8eb27defd14759e8522786061b13728623452740ba05055f5ba9d3d5}\\
\texttt{0xb9b50821ec5f01bb19ec75e0f22264fa9369436544b65c7cf653109dd26ef1f65c4fcaf1b1bcd2a7278afc34455d3da6}
\\
\texttt{0x965a05a1ba338f4bbbb97407d70659f4cea2146d83ac5da6c2f3de824713c927dcba706f35322d65764912e7756103e2}
}\\

\midrule

bloXroute (E) & \texttt{0xf573d99385c05c23b24ed33de616ad16a43a0919}& \specialcell{\texttt{0xb086acdd8da6a11c973b4b26d8c955addbae4506c78defbeb5d4e00c1266b802ff86ec7457c4c3c7c573fa1e64f7e9e0}\\\texttt{0x95701d3f0c49d7501b7494a7a4a08ce66aa9cc1f139dbd3eec409b9893ea213e01681e6b76f031122c6663b7d72a331b}
\\\texttt{0x82801ab0556f7df1fb9bb3a61ca84beea8285a8dc3c455a7ea16a8b2993fe06058e0e7d275b28ea5d9f2ae995aa72605}}\\

\midrule

blocknative &\texttt{0xbaf6dc2e647aeb6f510f9e318856a1bcd66c5e19} & \specialcell{\texttt{0x8000008a03ebae7d8ab2f66659bd719a698b2e74097d1e423df85e0d58571140527c15052a36c19878018aaebe8a6fea} \\
\texttt{0x9000009807ed12c1f08bf4e81c6da3ba8e3fc3d953898ce0102433094e5f22f21102ec057841fcb81978ed1ea0fa8246} \\
\texttt{0xa66f3abc04df65c16eb32151f2a92cb7921efdba4c25ab61b969a2af24b61508783ceb48175ef252ec9f82c6cdf8d8fd} \\
\texttt{0xa00000a975dffbd1ef61953ac6c90b52b70eb0188eb9d030774346c9248f81e875f7e8bc56c4bbbda297a9543cfa051d} } \\

\midrule

builder0x69 & \texttt{0x690b9a9e9aa1c9db991c7721a92d351db4fac990} &
\specialcell{\texttt{0x8bc8d110f8b5207e7edc407e8fa033937ddfe8d2c6f18c12a6171400eb6e04d49238ba2b0a95e633d15558e6a706fbe4} \\
\texttt{0xb194b2b8ec91a71c18f8483825234679299d146495a08db3bf3fb955e1d85a5fca77e88de93a74f4e32320fc922d3027} \\
\texttt{0xa971c4ee4ac5d47e0fb9e16be05981bfe51458f14c06b7a020304099c23d2d9952d4254cc50f291c385d15e7cae0cf9d} \\
\texttt{0xa4fb63c2ceeee73d1f1711fadf1c5357ac98cecb999d053be613f469a48f7416999a4da35dd60a7824478661399e6772} \\
\texttt{0xb8fceec09779ff758918a849bfe8ab43cea79f6a98320af0af5b030f6a7850fcc5883cb965d02efb10eed1ffa987e899}} \\

\midrule

Builder 1& \texttt{0x473780deaf4a2ac070bbba936b0cdefe7f267dfc} &\specialcell{\texttt{0xa1daf0ab37a9a204bc5925717f78a795fa2812f8fba8bda10b1b27c554bd7dedd46775106facd72be748eea336f514e9}\\\texttt{0x89783236c449f037b4ad7bae18cea35014187ec06e2daa016128e736739debeafc5fe8662a0613bc4ca528af5be83b3c}}\\

\midrule

Builder 2 & \texttt{0xbd3afb0bb76683ecb4225f9dbc91f998713c3b01} &\texttt{0x82ba7cadcdfc1b156ba2c48c1c627428ba917858e62c3a97d8f919510da23d0f11cf5db53cb92a5faf5de7d31bf38632}  \\

\midrule

Builder 3 & &\texttt{0xafc9274fe595e8cff421ab9e73b031f0dff707ea1852e2233ff070ef18e3876e25c44a9831c4b5f802653d4678ccc31f}  \\

\midrule

Builder 4 & \texttt{0x3b7faec3181114a99c243608bc822c5436441fff}&\texttt{0xa1f10d66aa4b73c5d9a6cc38a098b2c6ce031a6750ea2da01918ba3ac57c2ce1e39a0da622bd8ccd7c9930861f949fa2}  \\

\midrule

Builder 5 & \texttt{0xb646d87963da1fb9d192ddba775f24f33e857128}&\texttt{0x8bcd1148e83d0a844d2d42f90df0837dbe407055367b3bfcf04227e47ea65a0164fc13a66584aae286f8f7322dc69501}  \\

\midrule

Builder 6 & &\texttt{0xa25f5d5bd4f1956971bbd6e5a19e59c9b1422ca253587bbbb644645bd2067cc08fb854a231061f8c91f110254664e943}  \\

\midrule

Eden & \texttt{0xaab27b150451726ec7738aa1d0a94505c8729bd1} & \specialcell{\texttt{0x8e39849ceabc8710de49b2ca7053813de18b1c12d9ee22149dac4b90b634dd7e6d1e7d3c2b4df806ce32c6228eb70a8b}  \\
\texttt{0xa5eec32c40cc3737d643c24982c7f097354150aac1612d4089e2e8af44dbeefaec08a11c76bd57e7d58697ad8b2bbef5} \\
\texttt{0x91970c2db7c12510acb2e9c45844f7de602f83a7f31064f7ca04a807b607d7aebfc0abda73c036a92e5c3e56ebca04b7} \\
\texttt{0xa412007971217a42ca2ced9a90e7ca0ddfc922a1482ee6adf812c4a307e5fb7d6e668a7c86e53663ddd53c689aa3d350}   }\\

\midrule

eth-builder & \texttt{0xfeebabe6b0418ec13b30aadf129f5dcdd4f70cea}  &
\specialcell{\texttt{0x8eb772d96a747ba63af7acdf92dc775a859f76a77e4c6ed124dca6360e74e4e798a75a925eb8fd0dde866317fff18ad0} \\\texttt{0x8ea1393f49d894ae22ec86e38d9aeb64b8336dac947e69cb8468acf510d010ce0b51b21ac3e1244bdb91c52e020ea525} } \\

\midrule

Flashbots & \specialcell{\texttt{0xb64a30399f7F6b0C154c2E7Af0a3ec7B0A5b131a} \\
\texttt{0xdafea492d9c6733ae3d56b7ed1adb60692c98bc5}}    &
\specialcell{\texttt{0x81babeec8c9f2bb9c329fd8a3b176032fe0ab5f3b92a3f44d4575a231c7bd9c31d10b6328ef68ed1e8c02a3dbc8e80f9} \\
\texttt{0x81beef03aafd3dd33ffd7deb337407142c80fea2690e5b3190cfc01bde5753f28982a7857c96172a75a234cb7bcb994f}  \\
\texttt{0xa1dead01e65f0a0eee7b5170223f20c8f0cbf122eac3324d61afbdb33a8885ff8cab2ef514ac2c7698ae0d6289ef27fc} } \\

\midrule

Manta-builer & \texttt{0x5f927395213ee6b95de97bddcb1b2b1c0f16844f}   &\specialcell{\texttt{0xa0d0dbdf7b5eda08c921dee5da7c78c34c9685db3e39e81eb91da94af29eaa50f1468813c86503bf41b4b51bf772800e}  \\\texttt{0xb1b734b8dd42b4744dc98ea330c3d9da64b7afc050afed96875593c73937d530a773e35ddc4b480f9d2e1d5ba452a469}  \\\texttt{0xb5a688d26d7858b38c44f44568d68fb94f112fc834cd225d32dc52f0277c2007babc861f6f157a6fc6c1dc25bf409046} }\\

\midrule

rsync-builder & \texttt{0x1f9090aae28b8a3dceadf281b0f12828e676c326}  &\specialcell{\texttt{0x978a35c39c41aadbe35ea29712bccffb117cc6ebcad4d86ea463d712af1dc80131d0c650dc29ba29ef27c881f43bd587} \\\texttt{0x83d3495a2951065cf19c4d282afca0a635a39f6504bd76282ed0138fe28680ec60fa3fd149e6d27a94a7d90e7b1fb640}
\\\texttt{0x945fc51bf63613257792926c9155d7ae32db73155dc13bdfe61cd476f1fd2297b66601e8721b723cef11e4e6682e9d87}} \\

\bottomrule
\end{tabular}
\end{adjustbox}

    \caption{Builder name, address(es), and public key(s) for top 17, in terms of number of block built, builders.}
    \label{tab:builderaddress}
\end{table*}   
In Table~\ref{tab:builderaddress}, we provide a map from the builder name to their Ethereum fee recipient address and public key for the biggest 17 builders in terms of the number of blocks built. Notably, we do not have a fee recipient address for Builder 3 and Builder 6 as these builders always note down the proposer address as the fee recipient. Thus, we find no trace of these builders on the Ethereum blockchain. 

\section{Proposer-Builder Profit Shares}\label{app:profit}

\begin{figure}[H]
        \centering
        \includegraphics[scale =1]{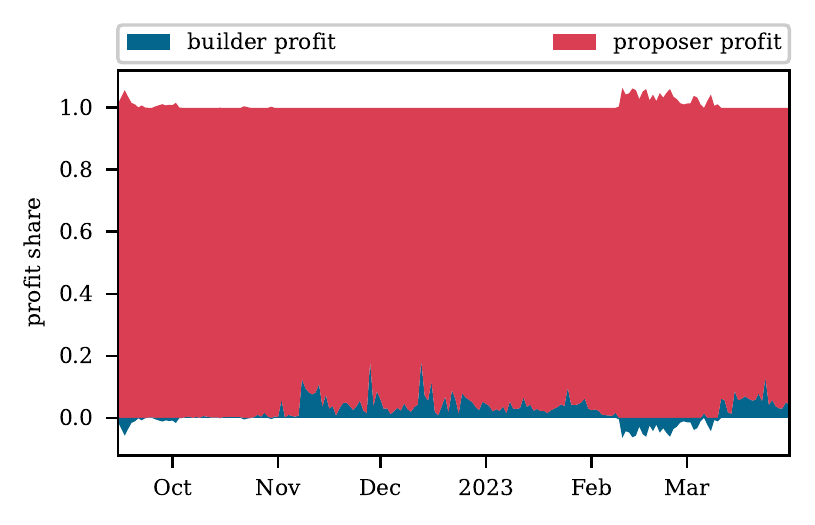}
        \caption{Daily profit share between block builder and block proposers.}
        \label{fig:feesplit}\vspace{-6pt}
    \end{figure}

    In Figure~\ref{fig:feesplit}, we explore the daily profit share between builders and proposers. We see that often the builders who subsidize blocks, i.e., have negative block profit, skew the daily average to negative. This behavior was initially prominent, then largely went away until mid-February when we saw an abnormal spike in negative builder profit. In investigating this further, we find that this negative average is almost entirely due to a single builder (beaverbuild), who took a 1.7K ETH loss this month. Though we cannot be certain of beaverbuild's behavior, we hypothesize that they developed another strategy to extract profit outside of the rewards we measure, and paid the proposers the competitive market price to build the blocks (the negative amount we see).

\balance
\section{MEV Transaction Types}\label{app:mev}

\begin{figure}[H]
        \centering
        \includegraphics[scale =1]{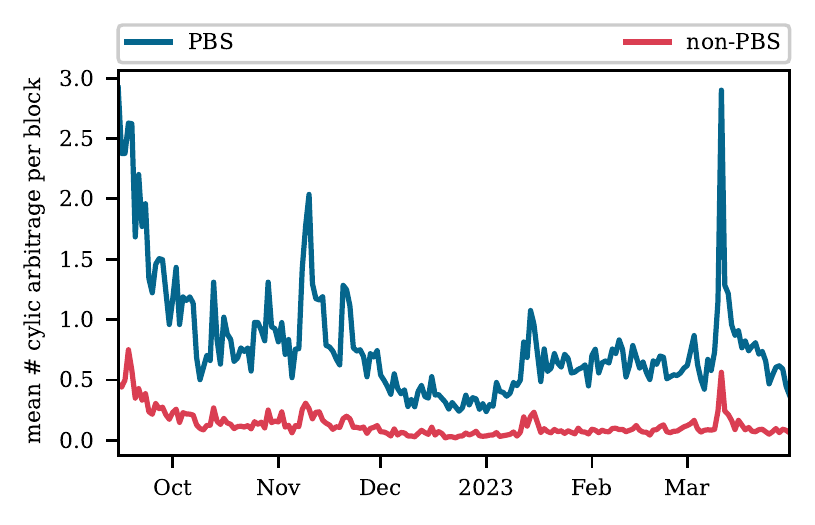}
        \caption{Daily mean number of arbitrage transactions per block for PBS (blue line) and non-PBS (red line) blocks.}
        \label{fig:arbitrage}\vspace{-6pt}
    \end{figure}

    In the following, we provide a more in-depth analysis concerning the number of various types of MEV transactions in PBS and non-PBS blocks. The most common type of MEV transaction we observe is sandwich attacks. In total, we observed 1,329,368 sandwich attacks during our data collection period. Note that each sandwich attack consists out of two transactions --- the front-running and the back-running transaction. We plot the daily average number of sandwich attacks in PBS and non-PBS blocks. There are almost no sandwich attacks in non-PBS blocks, while there is on average more than one sandwich attack in PBS blocks.     
    \begin{figure}[H]
        \centering
        \includegraphics[scale =1]{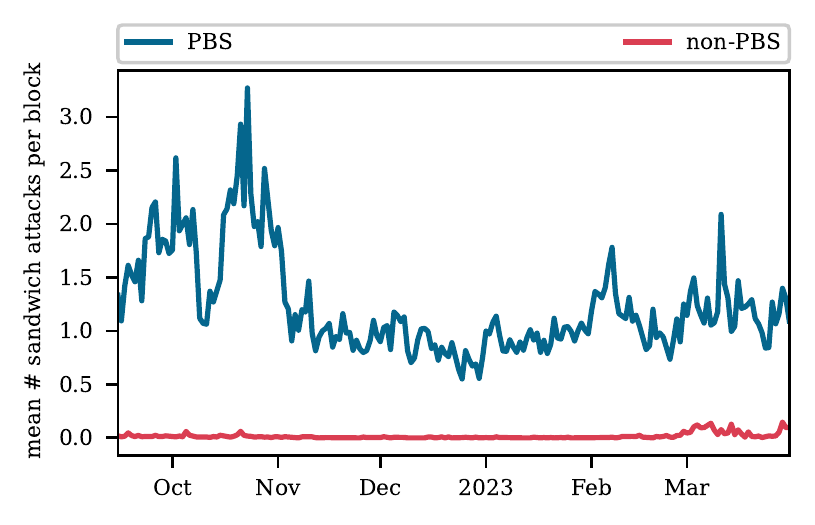}
        \caption{Daily mean number of sandwich attacks per block for PBS (blue line) and non-PBS (red line) blocks.}
        \label{fig:sandwich}\vspace{-6pt}
    \end{figure}
    We identify 871,560 cyclic arbitrage transactions. As we show in Figure~\ref{fig:arbitrage}, the vast majority of these are in PBS blocks. However, in comparison to sandwich attacks, the difference is less startling. While there are 0.72 cyclic arbitrage transactions in PBS blocks on average, there are 0.20 in non-PBS blocks. The difference appears biggest on days with total high numbers of cyclic arbitrage transactions. 
    The by far rarest type of MEV transaction we track is liquidations. In total, we identified 4,173 liquidations. Thus, the average number of liquidations per block is low in both PBS and non-PBS blocks. Further, the difference between PBS and non-PBS blocks is the smallest. While there are 0.003 liquidations per non-PBS block on average, there are 0.02 liquidations per PBS block on average. We believe that the smaller difference might be related to the time-sensitive nature of liquidations. A position of a lending protocol becomes available for liquidation once the price oracle updates. These updates can take place in PBS and non-PBS blocks. 
      \begin{figure}[H]
        \centering
        \includegraphics[scale =1]{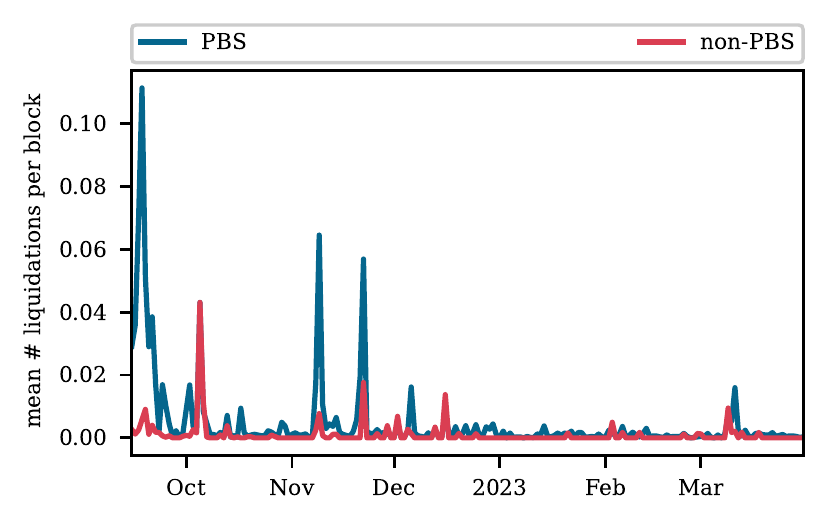}
        \caption{Daily mean number of liquidations per block for PBS (blue line) and non-PBS (red line) blocks.}
        \label{fig:liquidation}\vspace{-6pt}
    \end{figure}

\end{document}